\documentclass[twocolumn,trackchanges]{aastex701}

\usepackage{CJKutf8}
\usepackage{microtype}
\usepackage[T1]{fontenc}
\usepackage{amsmath}
\allowdisplaybreaks
\usepackage{multirow}
\usepackage{bm}
\usepackage{dsfont}
\usepackage{mathrsfs}
\usepackage[dvipsnames]{xcolor}
\usepackage{soul}
\usepackage[normalem]{ulem}
\usepackage{fontawesome}
\usepackage{graphicx}
\usepackage{booktabs}
\usepackage{placeins}
\usepackage{xspace}
\usepackage[figure, figure*]{hypcap}
\usepackage{todonotes}
  \setuptodonotes{
    backgroundcolor=lightgray!50,
    linecolor=lightgray,
    bordercolor=none,
    tickmarkheight=0.2em,
  }

\newcommand{\gkai}[1]{\begin{CJK*}{UTF8}{gkai}\raisebox{.1em}{(}#1\raisebox{.1em}{)}\end{CJK*}}

\bmdefine{\vx}{x}
\bmdefine{\vk}{k}
\bmdefine{\ve}{e}
\bmdefine{\vg}{g}
\bmdefine{\vm}{m}

\newcommand{\hatbf}[1]{\hat{\mathbf{#1}}}
\newcommand{\pard}[2]{\tfrac{\partial #1}{\partial #2}}

\newcommand{\NUMTRIALS}{\NUMTRIALSVAL\xspace}
\newcommand{\NUMTRIALSVAL}{{200}}    
\newcommand{\NUMRETRAIN}{{5}}        
\newcommand{\SNFINAL}{{0.2552\pm 0.0002}}  
\newcommand{\SNFPFS}{{0.3092 \pm 0.0004}}   
\newcommand{\SNREL}{{17.5}}              
\newcommand{\MONEFINAL}{{0.524 \times 10^{-3} \pm 0.407 \times 10^{-3}}}
\newcommand{\MTWOFINAL}{{1.326 \times 10^{-3} \pm 0.363 \times 10^{-3}}}
\newcommand{\SIGMASEL}{0.5}
\newcommand{\RHOMIN}{0.7}
\newcommand{\SLOPEMIN}{-0.01}

\newcommand{\SNEMHLR}{{0.2488 \pm 0.00020}}    
\newcommand{\SNEMHLRREL}{{19.6}}              
\newcommand{\MONEEMHLR}{1.52 \times 10^{-3} \pm 0.28 \times 10^{-3}}
\newcommand{\MTWOEMHLR}{1.51 \times 10^{-3} \pm 0.30 \times 10^{-3}}

\shorttitle{A Unified Statistical Framework for Weak Gravitational Lensing Shear Estimation}
\shortauthors{Lin et al.}

\begin{document}

\allowdisplaybreaks[4]

\title{Slay the Shear: \\
A Unified Statistical Framework for Weak Gravitational Lensing Shear Estimation}

\correspondingauthor{Shurui Lin}
\email{shuruil3@illinois.edu}

\author[orcid=0009-0000-5381-7039]{Shurui Lin\gkai{林书睿}}
\affiliation{Department of Astronomy, University of Illinois
Urbana-Champaign, 1002 West Green Street, Urbana, IL 61801, USA}
\email[]{shuruil3@illinois.edu}
\author[orcid=0000-0003-2880-5102,gname=Xiangchong, sname='Li']{Xiangchong Li}
\affiliation{Brookhaven National Laboratory, Bldg 510, Upton, New York 11973, USA}
\email[]{xli6@bnl.gov}
\author[orcid=0000-0003-0049-5210]{Xin Liu}
\affiliation{Department of Astronomy, University of Illinois Urbana-Champaign, 1002 West Green Street, Urbana, IL 61801, USA}
\affiliation{National Center for Supercomputing Applications, University of Illinois Urbana-Champaign, Urbana, IL 61801, USA}
\affiliation{Center for Artificial Intelligence Innovation, University of Illinois Urbana-Champaign, 1205 West Clark Street, Urbana, IL 61801, USA}
\email[]{xinliuxl@illinois.edu}

\begin{abstract}
Weak gravitational lensing shear measurements are fundamentally limited by shape noise arising from the intrinsic diversity of galaxy morphologies.
Upcoming surveys such as Rubin/LSST, Euclid, and Roman demand more flexible, statistically optimal approaches that can fully exploit high-dimensional image information.
In this work, we develop a unified theoretical framework for shear estimation that connects classical response-based methods, shape noise, and modern machine-learning estimators through the concept of the score function---the gradient of the image likelihood with respect to shear.
We show that, for a general spin-2 ellipticity definition, the ensemble shear response corresponds to an inner product between the estimator and the score function, and that the score provides the minimum-variance unbiased shear estimator.
By incorporating response into the classical inverse-variance weight, we prove that the response-weighted inverse-variance weight is a general shape-noise-minimizing weight, independent of the intrinsic shape distribution.
Furthermore, we propose Response-weighted Denoising Score Matching (RDSM) that exploits the remaining structure to reduce shape noise by ${\sim}\SNREL\%$ relative to moment-based methods at LSST 10-year depth while maintaining a multiplicative shear estimation bias below $2\times 10^{-3}$.
Our result clarifies the optimality of existing calibration techniques while revealing a principled pathway for constructing improved estimators via nonlinear shape transformations and learned representations.

\end{abstract}


\section{Introduction}\label{sec:intro}

Weak gravitational lensing is one of the most powerful probes of cosmology, enabling precise measurements of the matter distribution and the growth of structure across cosmic time \citep{Mellier1999,Bartelmann2001,Kilbinger2015,Mandelbaum2018}.
Current and upcoming Stage~IV experiments (Euclid, \citealp{EuclidCollaboration2024c}; Rubin/LSST, \citealp{Ivezic2019}; Roman, \citealp{Spergel2015}) will deliver galaxy shape measurements for billions of sources, requiring shear estimators that are not only unbiased at the sub-percent level but also near-optimal in variance \citep{Mandelbaum2018}.

Shear inference is fundamentally limited by shape noise: the irreducible variance from the intrinsic diversity of galaxy morphologies \citep{Mandelbaum2018}.
Traditional approaches mitigate this by calibrating estimator response via \textsc{Metacalibration} \citep{Huff2017,Sheldon2017} or analytic shear response models \citep{Li2023,Li2025_1,Li2025}, using fixed shape representations based on second moments or parametric fits \citep{KSB1995,Miller2013}.
These methods focus on achieving unbiasedness but do not explicitly optimize variance with respect to the full image likelihood.
Meanwhile, machine-learning (ML) approaches \citep{Tewes2019,Ribli2019a,Ribli2019b,Springer2020,Zhang2024} offer flexible, high-dimensional representations but often lack clear statistical interpretation or optimality guarantees \citep{Karniadakis2021}.
Recent work has proposed a D$_4$-equivariant CNN (D$_4$CNN) to parametrize the estimator $\hatbf{e}(I)$, and used AnaCal \citep{Li2023,Li2025} to compute the shear-response matrix $R_{ij} = \partial\langle\hat{e}_i\rangle/\partial g_j$ via back-propagation \citep{Lin2026}.
The challenge is therefore to find a statistically optimal way to harness the flexibility of ML.

Decades ago, early pioneers in weak lensing proposed an optimal weight for an ideal shape measurement by accounting for the intrinsic shape distribution \citep{Bartelmann2001,Bernstein2002}.
Their method has two basic starting points: analytical shear transformation, especially the M\"{o}bius transformation, for individual galaxies, and a known empirical model for the intrinsic ellipticity distribution.
As a result, it strongly relies on empirical model fitting when applied to real data.
Consequently, the inverse-variance weight, which is optimal only for a Gaussian intrinsic shape distribution with constant shear response, has become the standard choice in weak-lensing analyses, despite being suboptimal for the true non-Gaussian galaxy population.

To avoid the uncertainty and systematics from model fitting, data-driven methods are needed for practical implementation on real observations.
Exploiting the spin-2 properties of shear, the SYM-PDF method was proposed, where the shear is estimated by symmetrizing the distribution of measured ellipticity \citep{Zhang2016SYMPDF}.
However, a wealth of additional spin-0 image observables beyond galaxy ellipticity, such as magnitude, half-light radius, and the norm of higher-order morphological moments, carry information about galaxy properties that can help reduce shape noise.
This motivates the need for a unified theoretical framework that accommodates general spin-2 shape definitions and auxiliary variables, together with a practical numerical method to implement the optimal weighting scheme on real data.

In this work, we address this by recasting shear estimation in terms of the \emph{score function}, i.e., the gradient of the log-likelihood with respect to shear.
The score is the minimum-variance unbiased estimator of shear \citep{Rao1945,Cramer1946,Lehmann1998}; in the weak-lensing context, related Bayesian arguments were developed by \citet{Bernstein2002}. We show that the ensemble shear response equals an inner product between the estimator and the score: $R_{ij} = \mathbb{E}[\hat{e}_i s_j]$.
This identity connects response calibration directly to statistical optimality.
Shape noise measures the misalignment angle between a chosen shear estimator and the score, and nonlinear transformations of ellipticity can reduce this angle systematically.
Furthermore, we propose a machine-learning strategy, Response-weighted Denoising Score Matching (RDSM), by integrating the shape response into the denoising score matching framework, serving as a data-driven method to calculate the score function for any given shape catalog.

The rest of this paper is organized as follows.  \autoref{sec:score} develops the statistical framework connecting the image likelihood to the score function, the central identity expressing the shear response as an inner product of shape and score, and the relation between shape noise and score alignment.
\autoref{sec:ML} presents the Response-weighted Denoising Score Matching (RDSM) approach for learning the score from data.
\autoref{sec:validation} empirically validates the shape noise prediction using the D$_4$CNN of \citet{Lin2026}.  \autoref{sec:discussion} discusses implications, connections to prior work, and future directions.  Notation is summarized in \autoref{tab:notation}; detailed derivations are given in the appendices.

\begin{table}
\centering
\caption{
    Summary of notation used throughout this paper.
    \label{tab:notation}
}
\begin{tabular}{ll}
\hline
Symbol & Definition \\
\hline
$I$ & \parbox[t]{0.7\columnwidth}{Observed galaxy image (pixel vector)} \\
$I_0$ & \parbox[t]{0.7\columnwidth}{Unlensed intrinsic image} \\
$i\text{-mag}$ & \parbox[t]{0.7\columnwidth}{\textit{intrinsic} $i$-band magnitude}\\
$\tilde{g} = g_1 + ig_2$ & \parbox[t]{0.7\columnwidth}{Complex reduced shear}\\
$\mathbf{g} = (g_1, g_2)^\top$ & \parbox[t]{0.7\columnwidth}{Real 2-vector shear} \\
    $\tilde{\epsilon}$ & \parbox[t]{0.7\columnwidth}{Complex distortion, $\tilde{\epsilon} = \frac{1-p}{1+p}e^{2i\phi}$, where $p$ is the axis ratio and $\phi$ is the position angle of the galaxy.} \\
$\hatbf{e}(I)$ & \parbox[t]{0.7\columnwidth}{Shape (also serves as the shear estimator); real 2-vector $\hat{e}_1, \hat{e}_2$} \\
$s_g(I)$ & \parbox[t]{0.7\columnwidth}{Score function, $s(I) = \partial\ln L(I|g)/\partial g$} \\
$s_0(\mathbf{e})$ & \parbox[t]{0.7\columnwidth}{Score function projected onto ellipticity space at zero shear} \\
$\mathbf{R}$ & \parbox[t]{0.7\columnwidth}{Ensemble shear response matrix, $R_{ij} = \langle\partial\hat{e}_i/\partial g_j \rangle$} \\
$\hatbf{R}(I)$ & \parbox[t]{0.7\columnwidth}{Per-image response, $\hatbf{R}(I) = \partial\hatbf{e}(T_\mathbf{g}(I_0))/\partial\mathbf{g}\big|_{\mathbf{g}=0}$} \\
$U_\phi$ & \parbox[t]{0.7\columnwidth}{Image rotation operator by angle $\phi$; induces the spin-2 action $U_{2\phi}$ on $\hatbf{e}$} \\
$\mathbf{F}$ & \parbox[t]{0.7\columnwidth}{Fisher information matrix} \\
$D(I,\mathbf{g})$ & \parbox[t]{0.7\columnwidth}{Detection/selection weight} \\
$M$ & \parbox[t]{0.7\columnwidth}{Moment-space measurement vector} \\
$f(I|\mathbf{g})$ & \parbox[t]{0.7\columnwidth}{Single-galaxy likelihood under shear $\mathbf{g}$} \\
$f_0(I_0)$ & \parbox[t]{0.7\columnwidth}{Intrinsic (zero-shear) galaxy image distribution} \\
$L(I|\mathbf{g})$ & \parbox[t]{0.7\columnwidth}{Product likelihood over all galaxies, $\prod_i f(I_i;\mathbf{g})$} \\
$L_0(\mathbf{e})$ & \parbox[t]{0.7\columnwidth}{Marginal distribution of ellipticity at zero shear} \\
\hline
\end{tabular}
\end{table}

\section{Score Function as the Optimal Shear Estimator}
\label{sec:score}

\begin{figure*}[ht]
\centering
\includegraphics[width=\textwidth]{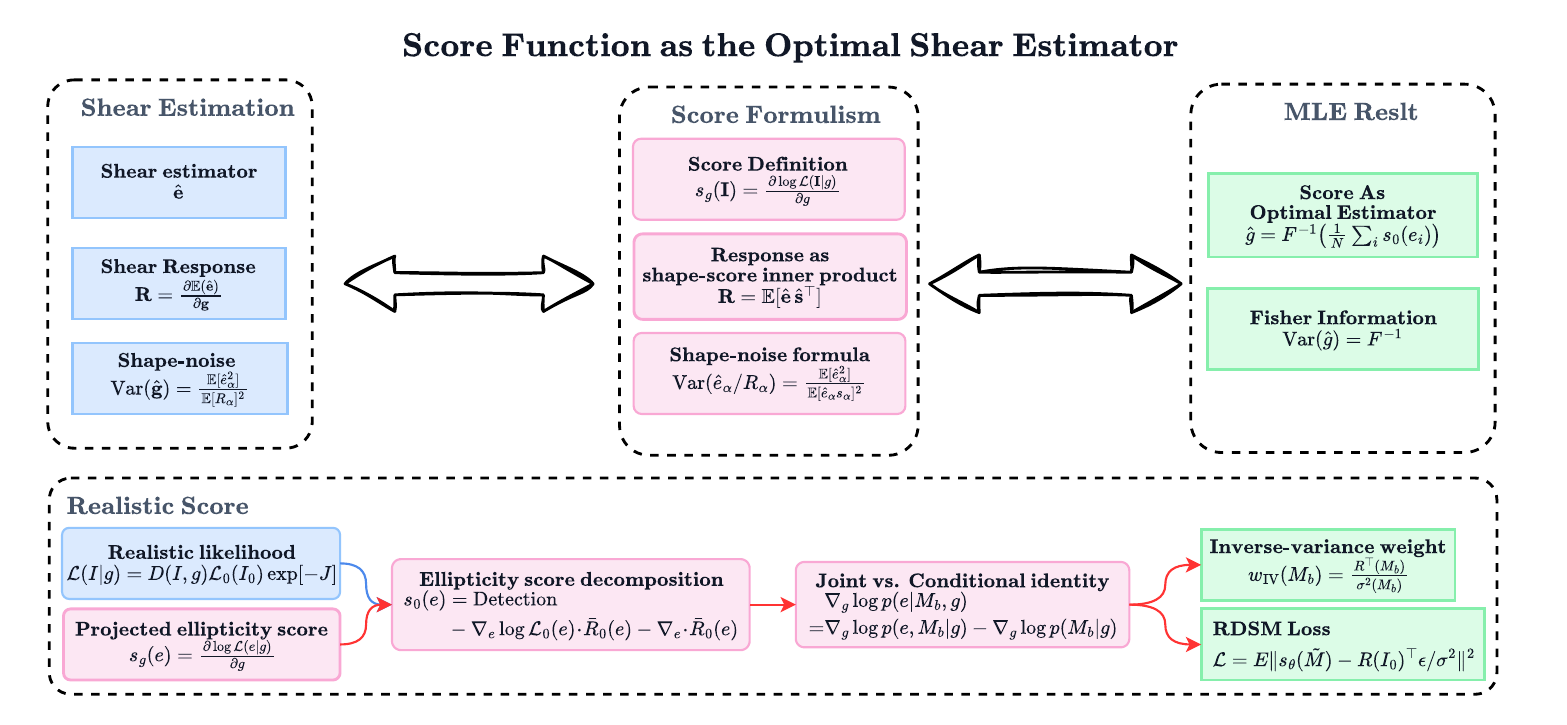}
\caption{
Schematic overview of the score-function framework (\autoref{sec:score}) and the RDSM estimator (\autoref{subsec:DSM}).
\textit{Top row:} the Fisher-score formulation (\autoref{sec:likelihood}) connects to response calibration (\autoref{sec:response_inner_product}) and to the interpretation of shape noise as misalignment with the score (\autoref{sec:shape_noise}).
    \textit{Bottom row:} Score function decomposition (\autoref{app:decomposition}), the joint/conditional score identity that links inverse-variance weighting (\autoref{sec:weight-hierarchy}) and the Response-weighted Denoising Score Matching method (\autoref{subsec:DSM}).
}
\label{fig:score_roadmap}
\end{figure*}

Since the shear is more than an order of magnitude smaller than the intrinsic galaxy shape in the weak-lensing regime, it can only be inferred at the statistical level.
The standard approach introduces a \textit{shear estimator}~$\hat{\mathbf{e}}(I)$, calibrates its expectation via the \textit{shear response matrix}~$\mathbf{R} = \partial\langle\hat{\mathbf{e}}\rangle/\partial\mathbf{g}$ \citep{Huff2017}, and characterizes the residual variance through the \textit{shape noise}~$\langle(\hat{\mathbf{e}})^2\rangle$.
This section shows that these three ingredients---estimator, response, and noise---are unified by a single object: the image score function $s(I) = \partial\ln L(I|\mathbf{g})/\partial\mathbf{g}$, the gradient of the log-likelihood with respect to shear.

Specifically, we first define the image likelihood and identify the score as the minimum-variance unbiased estimator of shear (\autoref{sec:likelihood}).
The central identity $R_{ij} = \mathbb{E}[\hat{e}_i \hat{s}_j]$ (\autoref{sec:response_inner_product}) follows directly from the likelihood structure, after which we show that the shape noise measures the misalignment angle between $\hat{\mathbf{e}}$ and $s$ in the $L^2(L_0)$ sense (\autoref{sec:shape_noise}).
We then project the image-space score onto the measurable shape to obtain a concrete optimal estimator (\autoref{sec:shape-projection}), and extend the projection to auxiliary flux and size observables, where the distinction between marginal and conditional scores arises (\autoref{sec:weight-hierarchy}).
\autoref{fig:score_roadmap} summarizes the overall structure.

\subsection{Likelihood Function of Images and Score Function}\label{sec:likelihood}
Shape measurement starts with pixelized galaxy images with random shear, denoted as $I$.
They are the result of the lensing effect (shear) on intrinsic galaxy images $I_0$, which follow a probability function $f_0(I_0)$, reflecting the intrinsic distribution of various galaxy properties and noise.
The lensed images and intrinsic images are related by $I = T_\mathbf{g}(I_0)$,
and this induces a transformation between the probability functions:
\begin{align}
    L(I|\mathbf{g}) = f_0[T_\mathbf{g}^{-1}(I)] \left| \det \pard{T_\mathbf{g}^{-1}(I)}{I} \right|.
\label{eq:fI_general}
\end{align}

Under the weak-lensing approximation $T_\mathbf{g}(I_0) \approx I_0 +\mathbf{g}\cdot\pard{I}{\mathbf{g}}|_{I_0}$, to first order in $g$:
\begin{equation}
  L(I|\mathbf{g}) = f_0\!\left(I - \mathbf{g}\,\frac{\partial T_\mathbf{g}^{-1}(I)}{\partial \mathbf{g}}\bigg|_g\right) \exp\!\left[-\mathbf{g} \cdot \,\mathrm{tr}\!\left(\frac{\partial^2 T_\mathbf{g}^{-1}(I)}{\partial I\,\partial \mathbf{g}}\right)\right].
\label{eq:parametrization}
\end{equation}
The exponential factor, often called the divergence or trace term, encodes the change in phase-space volume under shear and vanishes for area-preserving shear transformations at leading order in $\mathbf{g}$.

In real observations, we observe millions of galaxies, so we mostly
focus on the joint likelihood of multiple images given shear.
As intrinsic alignments of galaxies are usually separately modeled at the two-point level, we ignore the intrinsic alignment in this paper.
Throughout this paper we restrict to
the regime of isolated galaxies,
so that individual measurements can be treated
as statistically independent;
the more complex case in which galaxies overlap on the sky (blending), and the associated correlations between measurements,
lies beyond the scope of this work. Under this independence assumption, the
joint likelihood factorizes as:
\begin{align}
    L(\{I_i\}|\mathbf{g}) = \prod_i L(I_i|\mathbf{g}).
\end{align}
We simply denote this as $L(I|\mathbf{g})$ throughout the paper.

Under suitable regularity conditions and a flat prior, it can be
proven via Bayes' theorem that the maximum likelihood estimator of shear is proportional to the corresponding
\textit{score function} at zero shear (see \autoref{sec:bayesian} for details):
\begin{align}
    \hatbf{s}_\mathbf{0}(I) = \pard{\ln[L(I|\mathbf{0})]}{\mathbf{g}}.
    \label{eq:score_def}
\end{align}

And the estimator is:
\begin{align}
    \hatbf{g}_s = \mathbf{F}^{-1}\hatbf{s}(I),
\end{align}
where the \textit{Fisher information matrix} plays the role of response:
\begin{equation}
\mathbf{F}
\equiv
-\mathbb{E}\left[
\frac{\partial^2 \ln L(I\mid\mathbf{g})}{\partial \mathbf{g}\, \partial \mathbf{g}^\top}
\right]_{\mathbf{g}=\mathbf{0}}
=
\mathbb{E}\left[
\hatbf{s}_{\mathbf{0}}(I)\ \hatbf{s}_{\mathbf{0}}(I)^\top
\right].
\label{eq:fisher}
\end{equation}
And shape noise is equal to the inverse of the \textit{Fisher information matrix}:
\begin{equation}
\mathrm{Var}(\hatbf{g}_s) = \mathbf{F}^{-1}
\end{equation}

We emphasize that this optimality statement holds for a \emph{fixed}, known intrinsic galaxy population $f_0(I_0)$ under the weak-shear (linear) approximation; in practice, $f_0$ is unknown and must be inferred or marginalized over, so the score function derived here represents an idealized lower bound. Nevertheless, approximating the score function as closely as possible remains the principled target for any shear estimator, and we exploit this in \autoref{subsec:DSM}. In practice, the full image likelihood is intractable to evaluate directly from either observations or simulations, so a more practical approach is to construct an explicit unbiased estimator and seek to align it as closely as possible with the score function.

\subsection{Shear Response as the Inner Product of Shape and Score}
\label{sec:response_inner_product}
To align a shear estimator with the score function, we first establish how the two are related.
A common choice in practice is the linear estimator $\mathbf{R}^{-1}\hatbf{e}(I)$, with $\mathbf{R}^{-1}$ a constant matrix.
Because this estimator is unbiased to linear order in shear---in fact to second order, by spin-2 symmetry---we have:
\begin{align}
    \hat{\mathbf{g}} = \mathbb{E}\left(\mathbf{R}^{-1}\hatbf{e}\right) = \mathbf{g}.
\end{align}
Taking the gradient with respect to shear yields the response matrix:
\begin{align}
    \mathbf{R} = \pard{\mathbb{E}(\hatbf{e})}{\mathbf{g}}= \mathbb{E}_0\left(\pard{\hatbf{e}}{\mathbf{g}} \right),
    \label{eq:response}
\end{align}
where the subscript on $\mathbb{E}_0$ indicates that the average is taken over the intrinsic images $I_0$ at zero shear.
This identifies the constant matrix $R$ as the ensemble average of the per-image response at zero shear.
Writing the expectation as an integral over the likelihood $L(I\,|\,\mathbf{g})$, the gradient with respect to $\mathbf{g}$ can be moved onto the likelihood:
\begin{equation}
    R_{ij}
    = \pard{\mathbb{E}(\hat{e}_i)}{g_j}
    = \int \hat{e}_i\, \pard{L(I\,|\,\mathbf{g})}{g_j}\, \mathrm{d}I
    = \mathbb{E}\!\left[\hat{e}_i\, \pard{\ln L(I\,|\,\mathbf{g})}{g_j}\right].
\end{equation}
Recalling the definition of the score function in \autoref{eq:score_def}, the response takes the compact form
\begin{equation}
    R_{ij} = \mathbb{E}[\hat{e}_i\, \hat{s}_j],
    \label{eq:response_score}
\end{equation}
i.e., each entry $R_{ij}$ of the response matrix is the inner product (ensemble average) of the shape component $\hat{e}_i$ with the score component $\hat{s}_j$; a step-by-step reduction is given in \autoref{app:r_reduction}.

\subsection{Shape Noise as Misalignment between Shape and Score}
\label{sec:shape_noise}
Using the response formula of \autoref{eq:response_score}, we now derive the variance of the linear estimator $R^{-1}\hatbf{e}$ at zero shear, commonly known as the shape noise:
\begin{equation}
\begin{split}
    \mathrm{Var}(R^{-1}\hatbf{e})
    & = (R^{-1})^2\mathbb{E}(\hatbf{e}^2) \\
    & = [\mathbb{E}(\hatbf{e}\hatbf{s})^{-1}]^2\mathbb{E}(\hatbf{e}^2).
\end{split}
\end{equation}
A constant angle offset between shape and shear would introduce an off-diagonal response; we consider the aligned case, neglect the off-diagonal term, and treat each component $\hat{e}_i$ independently:
\begin{align}
    \mathrm{Var}\left(\frac{\hat{e_i}}{R_{i}}\right)
    = \frac{\mathbb{E}(\hat{e}_i^2)}{\mathbb{E}(\hat{e}_i\hat{s}_i)^2}
\label{eq:shape_noise_angle}
\end{align}
The Cram\'{e}r--Rao inequality \citep{Rao1945,Cramer1946} bounds the variance of any unbiased estimator by the inverse Fisher information,
\begin{equation}
\mathrm{Var}(\hat{\mathbf{g}}) \;\geq\; \mathbf{F}^{-1},
\label{eq:cramer-rao}
\end{equation}
with $\mathbf{F}$ defined in \autoref{eq:fisher}.
This bound has a geometric reading. Regard each shear component of the estimator $\hatbf{e}$ and of the score $\hatbf{s}$ as a function over the galaxy population, square-integrable under the zero-shear distribution $L_0$; such functions form the Hilbert space $L^{2}(L_0)$, with inner product $\langle f,g\rangle = \mathbb{E}_{0}[f\,g]$ given by the population average at zero shear. The angle $\Theta$ between estimator and score in this space,
\begin{equation}
\cos\Theta = \frac{\langle \hat{e}_i,\,\hat{s}_i\rangle}{\sqrt{\langle \hat{e}_i,\,\hat{e}_i\rangle\,\langle \hat{s}_i,\,\hat{s}_i\rangle}},
\label{eq:cos-align}
\end{equation}
recasts \autoref{eq:shape_noise_angle} as
\begin{equation}
\mathrm{Var}\!\left(\frac{\hat{e}_i}{R_i}\right)
= \frac{1}{\cos^{2}\Theta}\,\frac{1}{\mathbb{E}(\hat{s}_i^{2})}
= \frac{\mathbf{F}_{ii}^{-1}}{\cos^{2}\Theta},
\label{eq:shape_noise_cos}
\end{equation}
using $\mathbb{E}(\hat{s}_i^{2}) = \mathbf{F}_{ii}$. The shape noise is thus the Cram\'{e}r--Rao floor inflated by $1/\cos\Theta$, i.e.\ $\sigma_e = \sigma_e^{\min}/\cos\Theta$; since $\cos\Theta \le 1$ (Cauchy--Schwarz), the bound is saturated exactly when $\hatbf{e}/R$ is parallel to the score in the Hilbert space $L^{2}(L_0)$.

When $\hatbf{e}(I)$ is taken spin-2 covariant by construction---under an image rotation by angle $\phi$, $\hatbf{e}(U_\phi I) = U_{2\phi}\,\hatbf{e}(I)$, matching the transformation law of the score $\hatbf{s}(I)$---the two are parallel in direction at each image, i.e.\ as 2-vectors in image space. The residual misalignment in $L^{2}(L_0)$ is therefore purely radial: the remaining freedom in the estimator is a single scalar weight $h(I)$, with $\hatbf{s}(I)=h(I)\,\hatbf{e}(I)$. Saturating the bound then amounts to matching this weight $h(I)$ to that of the score across the population.

\subsection{Score Projected onto Shape}\label{sec:shape-projection}
We now turn this geometric optimality condition into a concrete estimator by projecting the score onto a measurable summary of the image, beginning with the shape.

Consider first the classical complex distortion $\tilde{\epsilon}\equiv[(a-b)/(a+b)]\,e^{2i\varphi}$ of a galaxy modeled as an ellipse with semi-axes $a\ge b$ and position angle $\varphi$, which is confined to the unit disk $|\tilde{\epsilon}|\le1$. A reduced shear $\tilde{g}$ acts on it as an isometry of this hyperbolic ``distortion disk'' \citep[][Eq.~2-13]{Bernstein2002}; to first order in shear this takes the M\"{o}bius form $\tilde{\epsilon}' \approx (\tilde{\epsilon} + \tilde{g})/(1 + \tilde{g}\cdot\tilde{\epsilon})$ \citep[][Eq.~3-31]{Bernstein2002}.
In this idealized description, each galaxy is fully specified by its distortion: the image $I$ is reduced to a single observable $\tilde{\epsilon}$, and the likelihood $L(\tilde{\epsilon}\mid\tilde{g})$ is defined directly in ellipticity space. The corresponding score reproduces the optimal estimator of \citet{Bernstein2002}.

Real galaxies, however, are not simple ellipses, so for a general image $\hatbf{e}(I)$ is a nontrivial function of the pixels rather than a fundamental parameter, and two obstacles prevent using the image-space score directly. First, the population likelihood $L(I\mid\mathbf{g})$ is not known in closed form, so $\hatbf{s}(I)$ is inaccessible. Second, even if it were, a usable estimator should depend on the image only through a low-dimensional, measurable summary. Both motivate \emph{projecting} the image-space score onto a chosen summary, which yields the optimal estimator within the class of estimators built from that summary. The minimal choice is the two-component shape $\hatbf{e}$, onto which the image-space score projects as (\autoref{app:e_space_param}):
\begin{equation}
   \hatbf{s}_\mathbf{g}(\mathbf{e}) = \int \mathrm{d}I \pard{\ln L(I|\mathbf{g})}{\mathbf{g}}
   \delta(\mathbf{e}-\hatbf{e}(I)) = \pard{\ln L(\mathbf{e}|\mathbf{g})}{\mathbf{g}}.
\end{equation}

The shape noise can likewise be expressed in $\mathbf{e}$ space (\autoref{app:sn_espace}):
\begin{equation}
\begin{split}
\mathrm{Var}(\hat{g}_a)
&= \frac{\mathbb{E}_{\mathbf{e}}\!\left[\hat{\mathcal{E}}_a^2(\hat{\mathbf{e}})\right]}
         {\mathbb{E}_{\mathbf{e}}\!\left[\hat{\mathcal{E}}_a(\hat{\mathbf{e}})\, \hat s_a(\hat{\mathbf{e}})\right]} \\
&\ge \frac{1}{\mathbb{E}_{\mathbf{e}}\!\left[\hat s_a^2(\hat{\mathbf{e}})\right]} ,
\end{split}
\end{equation}
where $\mathbb{E}_{\mathbf{e}}$ stands for expectation in the $\mathbf{e}$ space, and the minimum is reached when $\hat{\mathcal{E}} = \hatbf{s}(\hatbf{e})$.
Under small shear decomposition, it is possible to further decompose the score function into a linear combination of a detection term, a likelihood gradient term, and a response divergence term (see \autoref{app:decomposition}).

\subsection{Score Projected onto Shape, Flux, and Size}
\label{sec:weight-hierarchy}
In \autoref{sec:shape-projection}, projecting onto the two-component shape reduced the design of an optimal estimator to a single radial weight $h(I)$ acting on $\mathbf{e}$.
Projecting onto $\mathbf{e}$ alone, however, discards measurable information carried by other image features.
We recover part of it by enlarging the summary: alongside $\mathbf{e}$ we include additional spin-0 observables $m$ (flux, size, or higher-order moment combinations) carried by the image.
Extending the score formalism from $\mathbf{e}$ alone to the joint observable $(\mathbf{e}, m)$, two physically distinct definitions of ``the score'' arise---joint and conditional---leading to different estimators with different optimality properties.
This subsection sets up that distinction; the canonical inverse-variance weight $w_{\rm IV}$ emerges as a special case, and the choice between joint and conditional directly motivates the two RDSM variants tested in \autoref{sec:weight-decomp}.

Concretely, since the spin-2 direction of the estimator is already fixed by $\mathbf{e}$, an auxiliary spin-0 observable $m$ enters only through the radial weight and the overall normalization. We develop the two parametrizations---joint and conditional---in turn below.

\textbf{Case 1 --- joint score.}
The full joint likelihood $L(e, m \mid g)$ treats $m$ as a shear-dependent observable through its response $\partial m/\partial g$ (\autoref{eq:m00-shear-response}).
The corresponding score captures Fisher information from both the shape channel $e$ and the spin-0 channel $m$.
Using the Case 1 score requires response calibration that accounts for the induced shift in $m$, not only in $e$.

\textbf{Case 2 --- conditional score.}
Given any observable $M_b$ treated as a fixed conditioning variable, the conditional and joint scores are related by the identity
\begin{equation}
\nabla_g \log L(e \mid M_b, g) = \nabla_g \log L(e, M_b \mid g) - \nabla_g \log L(M_b \mid g).
\label{eq:conditional-identity}
\end{equation}
When $M_b$ is strictly shear-independent ($\nabla_g \log L(M_b \mid g) = 0$), the conditional score coincides with the joint score. When $M_b$ is weakly shear-dependent, the two agree up to a correction of order $\nabla_g \log L(M_b \mid g)$. For $M_b$ strongly shear-dependent (as is the case for $M_b = M_{00}$, the zero moment shapelet mode \citep{Refregier2003a,Refregier2003}, through \autoref{eq:m00-shear-response}), the conditional and joint scores differ at leading order.

To derive a closed-form expression for the Case~2 conditional score, we adopt a flexible parametric form for the conditional density $p(\hatbf{e}\mid\mathbf{g}, M_b)$.
Let $\mathbf{R}(M_b)$ denote the conditional shear response of $\hatbf{e}$ at fixed $M_b$, $\sigma(M_b)$ a per-$M_b$ scale parameter, and $q_{M_b}(\cdot)$ a non-negative radial profile function whose shape is allowed to depend on $M_b$. We assume the conditional takes the location-shift radial form
\begin{equation}
p(\hatbf{e} \mid \mathbf{g}, M_b) = Z(M_b)^{-1}\, q_{M_b}\!\left(\frac{\|\hatbf{e} - \mathbf{R}^\top(M_b)\, \mathbf{g}\|^{2}}{2\, \sigma^{2}(M_b)}\right),
\label{eq:radial-family}
\end{equation}
with normalization $Z(M_b)\equiv\int q_{M_b}\!\big(\|\hatbf{e}\|^{2}/[2\sigma^{2}(M_b)]\big)\,\mathrm{d}\hatbf{e}$ ensuring $\int p(\hatbf{e}\mid\mathbf{g},M_b)\,\mathrm{d}\hatbf{e}=1$. The Gaussian special case $q_{M_b}(u)=e^{-u}$ is $M_b$-independent and is discussed below; for general non-Gaussian tails the shape of $q_{M_b}$ is allowed to vary with $M_b$.
A direct calculation (\autoref{app:radial-family-score}) gives the exact conditional score at zero shear:
\begin{equation}
s^{\rm cond}(\hatbf{e}, M_b) = -\frac{q_{M_b}'(u)}{q_{M_b}(u)} \cdot \frac{\mathbf{R}^{\top}(M_b)\, \hatbf{e}}{\sigma^{2}(M_b)}, \quad u = \frac{\|e\|^{2}}{2\, \sigma^{2}(M_b)}.
\label{eq:cond-score}
\end{equation}

This leads to three important points:
\begin{itemize}
\item \textbf{Optimal weight for Gaussian.}
For $q_{M_b}(u)\propto e^{-u}$ (circular Gaussian conditional), $q'_{M_b}/q_{M_b}=-1$, so $s^{\mathrm{cond}}(\mathbf{e},M_b)=\mathbf{R}^{\top}(M_b)\,\mathbf{e}/\sigma^{2}(M_b)$ and the inverse-variance weight
\begin{equation}
w_{\rm IV}(M_b) = \frac{\mathbf{R}^{\top}(M_b)}{\sigma^{2}(M_b)}
\label{eq:w-IV-canonical}
\end{equation}
recovers the score exactly.
The globally optimal weight (in the larger class $w(\mathbf{e},M_b)$) is $\mathbf{e}$-dependent for non-Gaussian profiles, with the $\mathbf{e}$-dependence entering through $-q'_{M_b}/q_{M_b}$; this is what the full RDSM of \autoref{subsec:DSM} learns.

\item \textbf{Gauge freedom of $\sigma(M_b)$.}
The choice of the scale $\sigma(M_b)$ is not unique: a rescaling $\sigma\to\lambda(M_b)\,\sigma$, $q_{M_b}(u)\to q_{M_b}(u/\lambda^2)$ leaves the density unchanged.
The conditional score is invariant under this transformation, since the change in $1/\sigma^{2}$ is compensated by $-q'_{M_b}/q_{M_b}$.
Meanwhile, it is worth noting that it is possible for a good choice of $\sigma(M_b)$ to exist such that $q_{M_b}(u)$ has no dependence on $M_b$ (e.g., inverse-variance for a Gaussian radial profile).

\item \textbf{The role of response.}
The direction of the conditional score, $R^{\top}(M_b)\,\mathbf{e}/\sigma^{2}(M_b)$, factors into a gauge-invariant response part $R^{\top}(M_b)$ and a gauge-dependent scale part $1/\sigma^{2}(M_b)$. The response part is determined by how $\mathbf{e}$ responds to shear at fixed $M_b$ and is independent of the parametrization of the radial profile; consequently, within the restricted class $w(M_b)$, the optimal weight is proportional to $R(M_b)$ for \emph{any} radial conditional.
\end{itemize}

\paragraph{Choice of conditioning variable.}
The performance of a Case 2 estimator depends on how shear-sensitive the conditioning variable is through \autoref{eq:conditional-identity}.
Conditioning on galaxy size or magnitude is convenient because $\nabla_g \log L(M_b \mid g)$ is small for them and they are available from most catalogs; hence, the Case 2 estimator conditioned on $M_{00}$ approximates the true conditional score up to a small residual.
If we require our measurement to have spin-2 symmetry, an alternative conditioning on the orthogonal ellipticity component (e.g., estimating $g_1$ via $\pard{\ln L_0(e_1|e_2)}{g_1}$, as $\pard{\log L(e_2 \mid g_1)}{g_1}$ is small compared to $\pard{\ln L_0(e_1|e_2)}{g_1}$ at leading order in spin-2-symmetric populations) would realize the exact conditional score with no residual correction.

\paragraph{Consequence for estimator design.} Any improvement over $w_{\rm IV}$ must come from one of two channels:
\begin{itemize}
\item \textbf{Non-Gaussian Information:} knowledge of $q_{M_b}$ in $p(e \mid M_b)$ beyond the Gaussian assumption.
\item \textbf{Marginal channel (Case 1):} Fisher information from the shear response of whichever observable is used as $M_b$.
\end{itemize}

The full RDSM of \autoref{subsec:DSM}, can realize the above two cases, with empirical results shown in \autoref{sec:weight-decomp}.
Closed-form weights spanning these two channels---from the inverse-variance limit to the full shape-and-flux weight---are derived in \autoref{app:weight-hierarchy}, building on the ensemble-average channel decomposition of \autoref{app:phi-decomp}.

\section{Machine-Learning Approach for the Score Function}
\label{sec:ML}

\subsection{Response-weighted Denoising Score Matching (RDSM)}
\label{subsec:DSM}
The analysis of \autoref{sec:score} identifies the score function
\begin{equation}
\hat{s}_{0,a}(I) = \frac{\partial}{\partial g_a} \ln L(I \mid \mathbf{0}),
\end{equation}
as the minimum-variance unbiased shear estimator, and shows that its projection onto a reduced feature space provides a unifying target for a wide range of machine-learning shear estimators. This section develops a data-driven method to approximate this score directly from simulations, in the realistic setting where neither the intrinsic image distribution $f_0(I_0)$ nor the resulting image likelihood is available in closed form.

We work in the parameter space of a measurement vector $\mathbf{M}$ extracted from each lensed image,
\begin{align}
\mathbf{M} = \hat{\mathbf{M}}(T_\mathbf{g}(I_0)),
\end{align}
where $\hat{\mathbf{M}}$ is the measurement operator (e.g., FPFS moments or a D$_4$CNN output) and $\mathbf{g}$ is the applied shear.

For small shear around $\mathbf g = \mathbf 0$, we expand to first order:
\begin{align}
\hat{\mathbf M}
\approx
\hat{\mathbf M}(I_0)
+
\hatbf{R}(I_0)\,\mathbf g,
\end{align}
with the response defined as
\begin{align}
\hatbf{R}(I_0)
=
\left.
\pard{\hat{\mathbf{M}}(T_\mathbf{g}(I_0))}
{\mathbf g}
\right|_{\mathbf g=0}.
\end{align}

The likelihood of observing moments $\mathbf M$ given shear is
\begin{align}
L(\mathbf M \mid \mathbf g)
=
\int dI_0 \;
L_0(I_0)\,
\delta\!\left(
\mathbf{M}- \hat{\mathbf{M}}(T_\mathbf{g}(I_0))
\right).
\label{eq:lik-of-m}
\end{align}

Our target is the score function of this measurement-space likelihood, evaluated at zero shear:
\begin{align}
    \hatbf{s}_\mathbf{0}(\mathbf{M}) =
    \left.
    \pard{\ln L(\mathbf M \mid \mathbf g)}{\mathbf{g}}
    \right|_{\mathbf g=0}.
\end{align}

However, this gradient is intractable to evaluate directly. The likelihood $L(\mathbf{M}\mid\mathbf{g})$ is defined by a marginalization over the unknown intrinsic image distribution $f_0(I_0)$ (\autoref{eq:lik-of-m}), and the resulting density itself is not available in closed form. As a consequence, $\partial L/\partial\mathbf{g}$ cannot be obtained from straightforward Monte Carlo sampling: doing so would require either a parametric model of $L_0$ or a finite-difference scheme that perturbs the applied shear $\mathbf{g}$ across simulation runs, both of which introduce additional modeling or computational cost.
Denoising Score Matching \citep[DSM;][]{Vincent2011} sidesteps this difficulty by replacing the original measurement $\mathbf{M}$ with a smoothed surrogate $\tilde{\mathbf{M}}$ obtained by adding Gaussian noise:
\begin{align}
\tilde{\mathbf M}
=
\mathbf M + \boldsymbol\epsilon,
\qquad
\boldsymbol\epsilon \sim \mathcal N(\mathbf 0, \sigma^2 \mathbf I).
\end{align}

We then target the score function of the smoothed variable,
$\mathbf s(\tilde{\mathbf M})
=
\left.
\nabla_{\mathbf g}
\ln L(\tilde{\mathbf M} \mid \mathbf g)
\right|_{\mathbf g=0}$,
and recover the original score by taking the limit $\sigma \to 0$.
As shown in \autoref{app:rdsm_target}, the target satisfies:
\begin{align}
\mathbf s(\tilde{\mathbf M})
=
\mathbb E
\left[
\frac{\mathbf R(I_0)^{\!\top}
\boldsymbol\epsilon}
{\sigma^2}
\;\middle|\;
\tilde{\mathbf M}
\right].
\end{align}

Such an expectation can be realized by
training a network $s_\theta(\tilde{\mathbf M})$ with the following MSE loss:
\begin{align}
\mathcal L
=
\mathbb E
\left\|
s_\theta(\tilde{\mathbf M})
-
\frac{\mathbf R(I_0)^{\!\top}\boldsymbol\epsilon}{\sigma^2}
\right\|^2.
\label{eq:dsm_loss}
\end{align}
Because the MSE-optimal regressor learns the conditional mean of its target, training \autoref{eq:dsm_loss} to convergence yields $s_\theta(\tilde{\mathbf{M}}) = \mathbb{E}[\mathbf{R}(I_0)^{\top}\boldsymbol\epsilon/\sigma^{2}\mid\tilde{\mathbf{M}}] = s(\tilde{\mathbf{M}})$, recovering the score on the smoothed measurement (\citealt{Vincent2011}; cf.~\autoref{app:rdsm_target}).

The factor $\mathbf{R}(I_0)^{\top}$ appearing in \autoref{eq:dsm_loss} is what motivates the name \emph{response-weighted}, and it marks the one place where our construction departs from standard DSM. Ordinary DSM \citep{Vincent2011} regresses $\boldsymbol\epsilon/\sigma^{2}$ and thus recovers the gradient of $\log L(\tilde{\mathbf{M}})$ with respect to the noisy measurement $\tilde{\mathbf{M}}$ itself; our target is instead the gradient with respect to shear, $\partial_{\mathbf{g}}\log L(\tilde{\mathbf{M}}\mid\mathbf{g})|_{\mathbf{g}=0}$. 
The chain rule links the two: a shear perturbation $\delta\mathbf{g}$ displaces the measurement by $\mathbf{R}(I_0)\,\delta\mathbf{g}$, so contracting the noise direction $\boldsymbol\epsilon/\sigma^{2}$ with the per-image response $\mathbf{R}(I_0)^{\top}$ converts a displacement in moment space into the corresponding displacement in shear direction. The response therefore acts as a per-image calibration weight, and this same re-weighting is what makes the loss dimensionally consistent.

In practice, $\sigma$ should be chosen to be comparable to the intrinsic scatter of the moments $\mathbf M$, so that the noisy moments $\tilde{\mathbf M}$ span the support of the true distribution without over-smoothing the score landscape.
We take a data-driven choice in this paper by normalizing $\mathbf{M} \rightarrow \mathbf{M}/ \mathrm{std}(\mathbf M)$, and take $\sigma$ to be around the $0.5$ level.

The optimal solution satisfies
\begin{align}
s_\theta(\tilde{\mathbf M})
=
\mathbb E
\left[
\frac{\mathbf R(I_0)^{\!\top}\boldsymbol\epsilon}
{\sigma^2}
\;\middle|\;
\tilde{\mathbf M}
\right]
=
\left.
\nabla_{\mathbf g}
\ln L(\tilde{\mathbf M} \mid \mathbf g)
\right|_{\mathbf g=0}.
\end{align}
This shows that the score on the smoothed measurement space can be learned by minimizing \autoref{eq:dsm_loss}, a standard DSM objective augmented by the response factor $\mathbf{R}(I_0)$. The derivation above does not invoke the weak-shear expansion, so the training target itself is exact at finite $\sigma$; the weak-shear approximation enters only when we identify the learned score with the optimal estimator of \autoref{sec:shape_noise}.

\begin{figure*}[ht]
\centering
\includegraphics[width=0.9\textwidth]{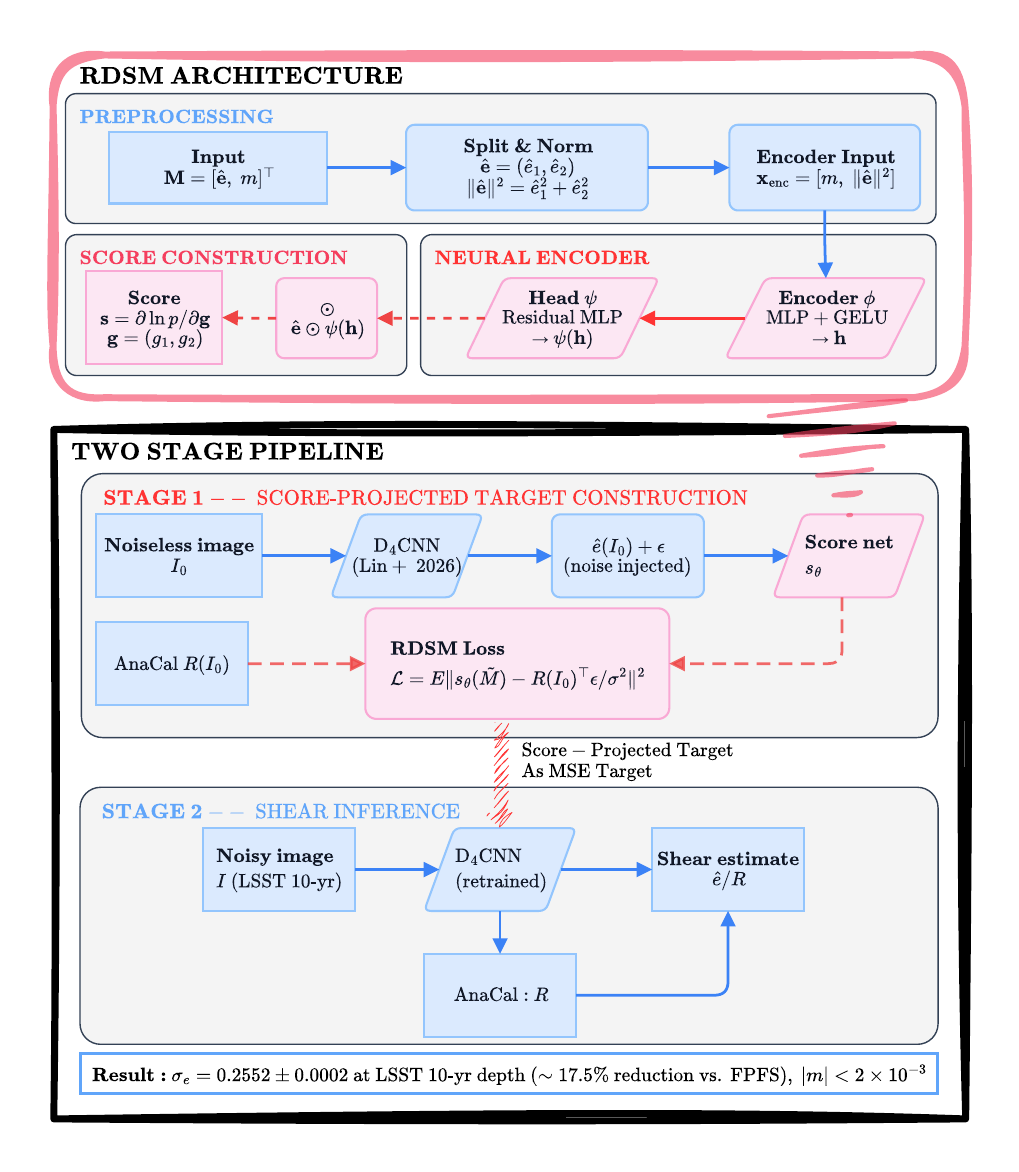}
\caption{
\textbf{Upper Panel:} Architecture for RDSM model.
The measurement $\mathbf{M}=[\hat{\mathbf{e}},\,m]^\top$ is separated into spin-2
($\hat{\mathbf{e}}$) and spin-0 ($m$) channels.
A neural encoder $\phi$ (MLP$+$GELU) processes the rotation-invariant combination
$\mathbf{x}_{\mathrm{enc}}=[m,\,\|\hat{\mathbf{e}}\|^2]$ into a hidden code
$\mathbf{h}$, which a residual MLP head $\psi$ decodes into spin-2 weights.
The shear score $\mathbf{s} = \hat{\mathbf{e}}\odot\psi(\mathbf{h})$ (with $\odot$ the
Hadamard product) is consequently parallel to $\hat{\mathbf{e}}$ on the spin-2 plane,
with the magnitude set by the encoder--decoder pair.\\
\textbf{Lower Panel:} Two-stage RDSM pipeline. \textit{Stage~1 (noiseless):} Controlled Gaussian noise $\boldsymbol\epsilon\sim\mathcal{N}(\mathbf{0},\sigma^2\mathbf{I})$ is injected into noiseless D$_4$CNN ellipticity predictions $\hat{\mathbf{e}}(I_0)$; a small network $s_\theta$ is trained with the RDSM loss of \autoref{eq:dsm_loss} using the AnaCal response $\mathbf{R}(I_0)$ as a calibration factor, producing score-projected labels $s_\theta(\tilde{\hat{\mathbf{e}}})$ that approximate the score in noiseless D$_4$CNN ellipticity space (\autoref{sec:training-target}). \textit{Stage~2 (noisy images):} The D$_4$CNN of \citet{Lin2026} is trained on noisy survey images using the Stage~1 labels as ground-truth training targets (replacing the FPFS ellipticity); AnaCal calibration then yields the response-normalized shear estimate $\hat{\mathbf{e}}/R$.
}
\label{fig:dsm_pipeline}
\end{figure*}

\subsection{Score-Projected Training Target}
\label{sec:training-target}

Following \autoref{subsec:DSM}, we construct the score-projected training labels in two stages, summarized in \autoref{fig:dsm_pipeline}.

\paragraph{Stage 1: noiseless score projection.}
We apply the RDSM framework of \autoref{subsec:DSM} to the noiseless D$_4$CNN ellipticity predictions $\hat{\mathbf{e}}(I_0)$, obtained by passing unlensed intrinsic galaxy images through the trained D$_4$CNN of \citet{Lin2026}.
The training set for RDSM contains D$_4$CNN shape and response for $1,000,000$ galaxies, where the per-image response $\mathbf{R}(I_0)$ is computed analytically via AnaCal \citep{Li2023,Li2025} as the shear response of the D$_4$CNN output.
Controlled Gaussian noise $\boldsymbol\epsilon \sim \mathcal{N}(\mathbf{0}, \sigma^2\mathbf{I})$ is injected, and a small network $s_\theta$ is trained with the loss of \autoref{eq:dsm_loss}. At convergence the optimal solution is the conditional expectation
\begin{equation*}
s_\theta(\tilde{\hat{\mathbf{e}}}) = \mathbb{E}\!\left[\mathbf{R}(I_0)^\top \boldsymbol\epsilon/\sigma^2 \,\big|\, \tilde{\hat{\mathbf{e}}}\right],
\end{equation*}
which approximates the score in noiseless D$_4$CNN ellipticity space.

\paragraph{Stage 2: supervised training on noisy images.}
The Stage~1 labels $s_\theta(\tilde{\hat{\mathbf{e}}})$ are then used as the ground-truth training target---in place of the FPFS ellipticity---when training the D$_4$CNN of \citet{Lin2026} on noisy observed images. The D$_4$CNN therefore learns the mapping from noisy pixels directly to score-aligned predictions in a single supervised pass. All other training settings (optimizer, learning rate, data augmentation, network architecture) are identical to \citet{Lin2026}, isolating the effect of the training-target change.

\paragraph{Role of spin-2 covariance.}
By design, the D$_4$CNN architecture is spin-2 covariant (\autoref{sec:model_architecture}), and the image-space score $s(I)$ shares this covariance. At each image, therefore, the score-projected target $s_\theta(\tilde{\hat{\mathbf{e}}})$ and the network output $\hatbf{e}(I)$ are parallel as 2-vectors on the spin-2 plane and can differ only in magnitude. The MSE loss of \autoref{eq:dsm_loss} thus acts purely on the radial (magnitude) profile and never on the angular direction; this is the precise sense in which RDSM matches $\hatbf{e}$ to the score in the Hilbert space $L^{2}(L_0)$ (\autoref{sec:shape_noise}). The spin-2 gauge that makes the response matrix $\mathbf{R}$ diagonal is also necessary for training stability: without it, the network output could rotate freely between epochs and the MSE loss would have no fixed minimum.

\subsection{Score Network Architecture}
\label{sec:model_architecture}

The architecture of the first-stage score network is shown in the upper panel of \autoref{fig:dsm_pipeline}. The model $s_{\boldsymbol\theta}(\mathbf{M})$ ingests a measurement vector
\begin{equation}
  \mathbf{M} = [\,\hat{\mathbf{e}},\; m,\,\ldots\,]^\top,
\end{equation}
comprising the D$_4$CNN ellipticity predictions $\hat{\mathbf{e}}$ on noiseless intrinsic images together with any additional spin-0 observables (flux, size, \ldots), and predicts the two-component shear score $\mathbf{s} = (s_1, s_2) = \partial \ln p(\mathbf{M}\mid\mathbf{g})/\partial\mathbf{g}$, where $\mathbf{g} = (g_1, g_2)$ is the gravitational shear.

\paragraph{Invariant encoder input.}
To enforce spin-2 covariance by construction, the encoder is fed only \emph{rotation-invariant} quantities. Defining $\| \hat{\mathbf{e}} \|^{2} = \hat{e}_1^{2} + \hat{e}_2^{2}$, the encoder input is
\begin{equation}
  \mathbf{x}_{\mathrm{enc}} = \bigl[\, m,\; \| \hat{\mathbf{e}} \|^{2} \,\bigr],
\end{equation}
which is unchanged under any rotation.

\paragraph{Encoder and decoder.}
The encoder $\phi$ is a standard multilayer perceptron (MLP)---a feedforward stack of fully connected linear layers separated by pointwise nonlinearities---that maps $\mathbf{x}_{\mathrm{enc}}$ to a hidden code $\mathbf{h}\in\mathbb{R}^{d_h}$. The width and depth of $\phi$ are treated as hyperparameters and tuned by the Optuna search of \autoref{sec:hyperparam_selection}; the activation function is the Gaussian Error Linear Unit \citep[GELU;][]{Hendrycks2016}, defined as
\begin{equation}
  \mathrm{GELU}(x) \;=\; x\,\Phi(x),
\end{equation}
where $\Phi$ is the standard normal CDF. The GELU is smooth (in contrast to ReLU's kink at the origin) and behaves like a soft gate: small inputs are nearly suppressed and large positive inputs are nearly passed through. We adopt it for its empirical training stability in regression settings, which we found important given that the RDSM target $\mathbf{R}(I_0)^{\top}\boldsymbol\epsilon/\sigma^{2}$ is itself a noisy quantity.

The hidden code $\mathbf{h}$ is then decoded by a residual MLP head $\psi$ into a 2-vector of spin-2 weights, and the final shear score is obtained by element-wise multiplication with the raw ellipticity,
\begin{equation}
  \mathbf{s} \;=\; \hat{\mathbf{e}}\,\odot\,\psi(\mathbf{h}),
\end{equation}

where $\odot$ denotes the Hadamard product. This multiplicative coupling enforces the output to be parallel to $\hat{\mathbf{e}}$ on the spin-2 plane while the magnitude (and the relative weighting of $\hat e_1$ vs.\ $\hat e_2$) is set by the encoder--decoder pair.

\paragraph{Training and stabilization.}
The score network is trained with the RDSM loss of \autoref{eq:dsm_loss}, with the response factor $\mathbf{R}(I_0)$ computed analytically by AnaCal \citep{Li2023,Li2025}.
To stabilize the early-stage training where the DSM target $\mathbf{R}(I_0)^{\top}\boldsymbol\epsilon/\sigma^{2}$ is dominated by noise, we project the target onto the direction of $\hat{\mathbf{e}}$ and apply a linear warmup that gradually blends the projected DSM target with a unit-vector proxy aligned with $\hat{\mathbf{e}}$.
This warmup discourages spurious early-epoch rotations of the predicted direction and converges to the pure DSM target by the end of warmup.

\subsection{Hyperparameter Selection}
\label{sec:hyperparam_selection}
The RDSM noise scale $\sigma$ and the network hyperparameters were selected by an Optuna \citep{Akiba2019} multi-objective study targeting joint minimization of multiplicative bias $|m|$ and shape noise $\sigma_e$, with the search space summarized in \autoref{tab:optuna}.
Four RDSM noise levels were evaluated in separate per-target runs. We then imposed two monotonicity constraints on the trained models: a Spearman rank correlation $\rho > \RHOMIN$ between $|s_\theta|$ and $|\hat{\mathbf{e}}|$, and a power-law slope satisfying $\mathrm{slope}-1 > \SLOPEMIN$, both measured on a held-out catalog.
These constraints reject hyperparameter combinations that produce significant non-monotone radial transformations in the bulk regime, where strong non-linearity is introduced (\autoref{sec:training-target}).
On the unconstrained Pareto front, all three values of $\sigma$ trace nearly identical bias-vs-shape-noise curves. The monotonicity constraints discriminate strongly: $\sigma = 0.3\,\mathrm{std}(\hat{e})$ trials fail the Spearman cut on most operating points (median $\rho_{\rm Spearman}\approx 0.41$ vs.~${\approx}\,0.90$ at $\sigma=0.5$), indicating non-monotone radial transformations that cause cross-component leakage between $g_1$ and $g_2$. Under the constraints, $\sigma = 0.5\,\mathrm{std}(\hat{e})$ dominates the front (\autoref{fig:pareto}), and we adopt it for the headline result.
Reproducibility was verified by retraining the chosen hyperparameter set $\NUMRETRAIN$ times with independent random seeds; the across-seed standard deviation in $|m|$ is reported alongside the central value.

\begin{deluxetable}{ll}
\tabletypesize{\footnotesize}
\tablecaption{Optuna hyperparameter search space and fixed settings.\label{tab:optuna}}
\tablehead{\colhead{Parameter} & \colhead{Value / Range}}
\startdata
Batch size & 5{,}000--200{,}000 (log-uniform) \\
Learning rate & $10^{-4}$--$(3\times10^{-3})$ (log-uniform) \\
Weight decay & $10^{-7}$--$10^{-3}$ (log-uniform) \\
Feature width & $\{64, 128, 256, 384, 512, 1024, 2048\}$ \\
Network depth & $[1, 4]$ \\
Head depth & $[2, 8]$ \\
Training epochs & 50 (fixed) \\
Noise injection target & $\{e_1\}$ or $\{e_1, e_2\}$ \\
DSM noise level $\sigma$ & $\{0.3, 0.4, 0.5, 0.6\}\,\mathrm{std}(\hat{\mathbf{e}})$ \\
Monotonicity: Spearman $\rho$ & $> \RHOMIN$ \\
Monotonicity: slope${}-1$ & $> \SLOPEMIN$ \\
Reproducibility retrains & $\NUMRETRAIN$ \\
\enddata
\tablecomments{All hyperparameters except training epochs were optimized via Optuna \citep{Akiba2019} with 50 epochs fixed. The monotonicity constraints reject non-monotone radial transformations that cause cross-component calibration leakage (\autoref{sec:training-target}).}
\end{deluxetable}

\begin{figure*}
\includegraphics[width=\textwidth]{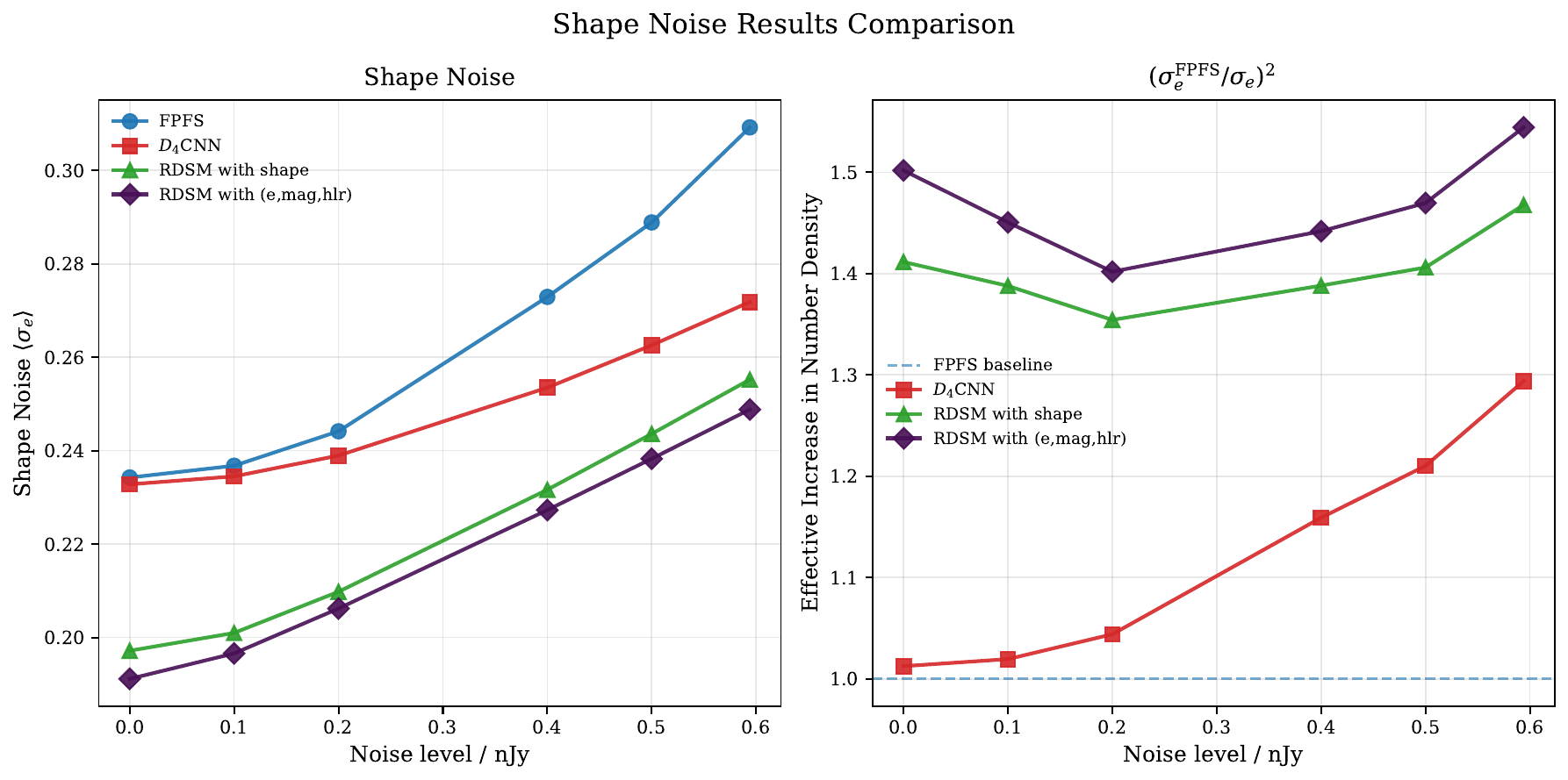}
\caption{
\textit{Left:} Shape noise $\sigma_e$ as a function of image noise level.
\textit{Right:} Effective increase in number density, $(\sigma_e^{\rm FPFS}/\sigma_e)^2$, relative to the FPFS baseline (dashed gray line at unity).
The four D$_4$CNN-based estimators are compared against the FPFS+AnaCal baseline (blue circles).
The D$_4$CNN trained with the standard FPFS ellipticity target \citep{Lin2026} (red squares) already reduces shape noise below FPFS.
RDSM with shape (green triangles) yields a further reduction, reaching $\SNFINAL$ at LSST 10-year depth, a $\sim\SNREL\%$ improvement over FPFS while satisfying $|m|{<}2\times10^{-3}$ (the headline result).
An exploratory Case~1 variant conditioned on the \textit{intrinsic} $i$-band magnitude and half-light radius, $(i\text{-mag},\,r_{1/2})$, at DSM noise level $\sigma = 0.6\,\mathrm{std}(\hat{\mathbf{e}})$ (dark purple pentagons) achieves lower shape noise than the headline but with calibration bias up to $|m|<2 \times 10^{-3}$.
The rightmost point on each curve corresponds to the LSST 10-year coadded $i$-band depth.
\label{fig:shape_noise_dsm}}
\end{figure*}

\section{Empirical Tests of the Score-Function Framework}
\label{sec:validation}

The shape noise bound derived in \autoref{sec:shape_noise} predicts that any estimator whose training target is better aligned with the score function will achieve lower shape noise, with the minimum reached when the target equals the projected score $s_0(\mathbf{e})$ (\autoref{eq:shape_noise_angle}).
We test this prediction through three complementary analyses in this section. \autoref{sec:sim} describes the LSST mock-image simulation infrastructure used for training and calibration.
\autoref{sec:validation_results} presents the shape noise improvement at LSST 10-year depth when the D$_4$CNN is trained with score-projected labels derived from the RDSM framework of \autoref{subsec:DSM}, including an exploratory Case~1 variant conditioned on intrinsic magnitude and half-light radius that extracts additional Fisher information from the marginal channel. Finally, \autoref{sec:weight-decomp} provides a noiseless decomposition of the total variance reduction into the optimal inverse-variance weight and the non-Gaussian channel, quantifying the information captured at each stage of the score-based estimator hierarchy.

\subsection{Simulation Setup}
\label{sec:sim}

We use the simulation infrastructure of \citet{Lin2026} unchanged: the
\texttt{xlens} package \citep{Li2025} with \textsc{GalSim} \citep{Rowe2015}, to
generate realistic, calibrated exposure images consistent with the LSST Science
Pipelines. With an input Data Challenge 1 (DC1) catalog of galaxies
\citep{Sanchez2020}, we generate images with $0.2\ \mathit{arcsec}$ pixel size,
convolved with Moffat PSF \citep{Trujillo2001Moffat} with FWHM$=0.8\
\mathit{arcsec}$. The noise in each pixel is drawn independently from a
Gaussian distribution with standard deviation equal to $0.594\ \mathrm{nJy}$ to
mimic the noise level of the LSST Y10 i-band coadd image. Each galaxy is placed
near the center of a $64 \times 64$ pixel postage stamp, after applying random
rotations, shear distortions, subpixel offsets, and PSF convolution. The
training set of D$_4$CNN contains $10,000$ simulated images with $i\text{-mag} {<} 24.5$
magnitude cut, and calibration tests are done with $40$ million isolated galaxies
for each sheared configuration. These settings match \citet{Lin2026} exactly.

\subsection{Shape noise reduction with RDSM}
\label{sec:validation_results}

\begin{figure}
\includegraphics[width=0.45\textwidth]{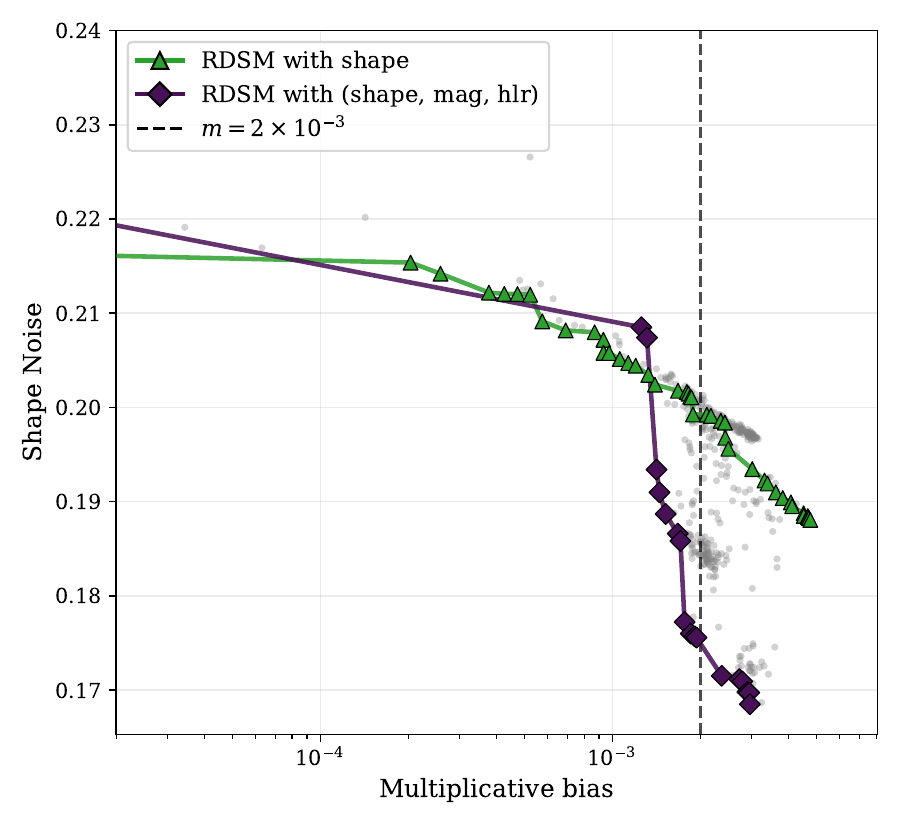}
\caption{
Bias--shape-noise Pareto front from Optuna multi-objective optimization of the RDSM hyperparameters ($\NUMTRIALS$ trials, search space defined in \autoref{sec:training-target}), under the monotonicity constraint $\rho_{\rm Spearman}(s_\theta, |e|) > \RHOMIN$ and shape-fit slope satisfying $\mathrm{slope}-1 > \SLOPEMIN$.
    Markers distinguish the two RDSM results: green triangles for shape-only results  with $\sigma=\SIGMASEL$ and purple pentagons for (shape, $i\text{-mag}$, $r_{1/2}$) result with $\sigma=0.6$ (in units of $\mathrm{std}(\hat{\mathbf{e}})$). The vertical dashed line at $|m| = 2 \times 10^{-3}$ indicates the LSST DESC requirement.
The operating point selected for \autoref{fig:shape_noise_dsm} lies at the intersection of the constrained front with this requirement, achieving shape noise $\SNFINAL$ at $|m_1| = \MONEFINAL$.
}
\label{fig:pareto}
\end{figure}

\autoref{fig:shape_noise_dsm} shows the shape noise as a function of image noise level for three estimators: (i) FPFS+AnaCal (baseline, blue circles), (ii) D$_4$CNN trained with the FPFS ellipticity target \citep{Lin2026} (red squares), and (iii) D$_4$CNN trained with the score-projected target derived here (green triangles and purple pentagons). 
At the LSST 10-year noise level, the score-projected target yields shape noise $\SNFINAL$, a $\sim\SNREL\%$ reduction relative to the FPFS baseline $\SNFPFS$.
The headline values are computed as the mean over $\NUMRETRAIN$ independent retrains at fixed hyperparameters; the quoted uncertainty is the across-retrain standard deviation.
Multiplicative biases at the selected operating point are $m_1 = \MONEFINAL$ and $m_2 = \MTWOFINAL$, both consistent with $|m| < 2 \times 10^{-3}$ at the $1\sigma$ level and confirming that the improved training target does not introduce significant additional calibration bias.
\autoref{fig:pareto} shows the constrained Pareto front from the Optuna multi-objective study; the operating point used here lies on this front under the monotonicity selection described in \autoref{sec:training-target}.

This result is consistent with the theoretical prediction of \autoref{sec:shape_noise}: applying RDSM to the D$_4$CNN output on noiseless images brings the effective training target closer to the score function.
Retraining the D$_4$CNN with score-aligned targets combines the denoising power of the neural network \citep{Lin2026} with statistical optimality, and thus reduces the shape noise on noisy images.
The theoretical connection to the Cram\'{e}r--Rao bound holds rigorously in the noiseless regime; the improvement observed here in the noisy regime is empirical evidence that score-aligned supervision transfers effectively through the D$_4$CNN.

As an empirical test of combining ellipticity with other spin-0 measurements, we trained a Case 1 RDSM variant conditioned on $(i\text{-mag},\,r_{1/2})$ in addition to $\mathbf{e}$, where $i\text{-mag}$ is the \textit{intrinsic} $i$-band magnitude in the simulation catalog and $r_{1/2}$ is the \textit{intrinsic} half-light radius.
Because both are intrinsic quantities, neither carries a shear response.
Using the intrinsic catalog values is an idealization, since the measured photometry and morphology in a real survey would be noisy and shear-dependent.
The Case 1 gain reported here should therefore be interpreted as a best-case demonstration of the marginal channel, and the residual improvement after propagating realistic measurement errors in magnitude and $r_{1/2}$ is left to future work.
This variant achieves shape noise $\SNEMHLR$ at LSST 10-year depth (purple pentagons in \autoref{fig:shape_noise_dsm}), a further $\sim\SNEMHLRREL\%$ reduction relative to the ellipticity headline (FPFS result, blue circles in \autoref{fig:shape_noise_dsm}), confirming that adding a second shear-sensitive observable extracts additional Case 1 Fisher information beyond what shape alone provides.
The multiplicative biases are $m_1 = \MONEEMHLR$ and $m_2 = \MTWOEMHLR$, still within the Stage-IV threshold $|m| <2 \times 10^{-3}$.
This demonstrates the effect of additional spin-0 components on reducing the shape noise.

\subsection{Empirical Decomposition of the RDSM Improvement}
\label{sec:weight-decomp}

The case analysis of \autoref{sec:weight-hierarchy} identifies two channels by which an estimator can improve on the inverse-variance weight $w_{\rm IV}$.
We now quantify the additional shape-noise reduction that RDSM achieves beyond $w_{\rm IV}$, denoted as the $w_{\rm IV} \to$ RDSM gap, on the \textit{noiseless} DC1 catalog ($i\text{-mag} < 24.5$, $10^6$ galaxies), thereby isolating the non-Gaussian information channel from pixel-noise effects.
Because this test uses noiseless images, the result is an idealized upper bound on the improvement from non-Gaussian structure, and the gain under realistic pixel noise (reported in \autoref{sec:validation_results}) is expected to be smaller.

For each galaxy we compute (i) FPFS moments and a power-law fit of shape variance $\sigma_g$ at different values of zero-order shapelet mode $m_{00}$:
$\sigma_g^2(m_{00}+C) = A(m_{00}+C)^\alpha$ (details can be found in \autoref{app:FPFS}; the zero-point shift constant $C$ is chosen to be $8.7$ in this paper);
(ii) the analytic response $R$ from AnaCal \citep{Li2023,Li2025};
(iii) the RDSM score from the trained D$_4$CNN of \autoref{sec:training-target}. \autoref{tab:weight-decomp} reports shape noise for each scheme.

\begin{table}
\centering
\footnotesize
\caption{Shape noise by estimator on noiseless DC1, $i\text{-mag} < 24.5$.
Three estimators are compared:
(i) FPFS (uniform weighting), serving as the noiseless reference before score-based optimization;
(ii) $w_{\rm IV}$, the optimal $e$-independent scaling weight (\autoref{eq:w-IV-canonical}),
(iii) Full RDSM , which learns the full shape distribution on top of $w_{\rm IV}$ through non-Gaussian structure in $p(\mathbf{e}\mid M_{00})$.
\label{tab:weight-decomp}}
\begin{tabular}{lc}
\hline
Estimator & $\sigma_e$ \\
\hline
\parbox[t]{0.45\columnwidth}{FPFS} & $0.2402{\pm}0.0004$  \\
\parbox[t]{0.45\columnwidth}{$w_{\rm IV}$} & $0.2156{\pm}0.0006$ \\
\parbox[t]{0.45\columnwidth}{Full RDSM with ellipticity and $M_{00}$} & $0.2021{\pm}0.0004$  \\
\hline
\end{tabular}

\smallskip
\footnotesize $\sigma_e$ is the root-mean-square of shape noise over $g_1, g_2$.
\end{table}

The total variance reduction from FPFS to RDSM is $\Delta\sigma_e^2 = \sigma^2_{\rm FPFS} - \sigma^2_{\rm RDSM} = 0.01685$, a $\sim 29\%$ decrease in $\sigma_e^2$ (or $\sim 16\%$ in $\sigma_e$). This reduction splits into two contributions:
\begin{align}
\Delta^2_{\rm scaling}
  &\equiv \sigma^2_{\rm FPFS} - \sigma^2_{\rm IV} = 0.01121 \quad (67\%), \label{eq:Delta-scaling}\\
\Delta^2_{\rm non\text{-}Gaussian}
  &\equiv \sigma^2_{\rm IV} - \sigma^2_{\rm RDSM} = 0.00564 \quad (33\%). \label{eq:Delta-nonGaussian}
\end{align}
The inverse-variance result (${\sim}67\%$) is already captured by the optimal $e$-independent weight $w_{\rm IV}$, which we proved in \autoref{sec:weight-hierarchy} to be the general scaling weight for \emph{any} radial conditional $p(\mathbf{e}\mid M_b)$.
RDSM achieves the additional reduction (${\sim}33\%$) by learning the actual intrinsic shape distribution, specifically the non-Gaussian structure in $p(\mathbf{e}\mid M_{00})$ that constitutes the first improvement channel identified in \autoref{sec:weight-hierarchy}.
Our test shows very little difference between the joint distribution $p(\mathbf{e},M_{00})$ and the conditional distribution $p(\mathbf{e}\mid M_{00})$ (i.e.\ Case~1 and Case~2 as defined in \autoref{sec:weight-hierarchy}), mostly because the measured $M_{00}$ has a small shear response; consequently the marginal (Case~1) channel contributes negligibly in the noiseless regime.
Pixel noise convolves $p(e \mid M_{00})$ with a Gaussian kernel of width $\propto \sigma_{\rm pix}/\mathrm{SNR}$, pushing the shape distribution toward a Gaussian form.
The non-Gaussian fraction reported here is therefore an upper bound on the RDSM advantage at LSST 10-year depth. A depth-dependent measurement of the split is deferred to future work.

These three estimators, summarized in \autoref{tab:weight-decomp}, trace the progression from uniform weighting to the full score-based optimum, with the $\Delta\sigma_e^2$ decomposition (\autoref{eq:Delta-scaling}--\autoref{eq:Delta-nonGaussian}) isolating the component captured by each improvement channel.

\section{Discussion and Future Work}
\label{sec:discussion}

\subsection{Summary of Main Results}

The central result of this paper is \autoref{eq:response_score}: the ensemble shear response equals the inner product between the estimator and the score function, $R_{ij} = \mathbb{E}[\hat{e}_i\,\hat{s}_j]$. This identity, which holds for any unbiased linear estimator satisfying the spin-2 condition of ellipticity definition, provides a precise optimality criterion: an estimator achieves the Cram\'{e}r--Rao bound if and only if it is proportional to the score. Three consequences follow directly.

\begin{enumerate}
\item \textbf{Shape noise and score misalignment.} Estimators poorly aligned with $s_\mathbf{g}(I)$ incur an avoidable penalty (\autoref{eq:shape_noise_angle}). The suboptimality of moment-based methods like FPFS at low S/N arises because their implicit second-moment weighting is not the score-optimal combination of pixel values; the $\sim 10\%$ improvement of D$_4$CNN$\times$AnaCal over FPFS \citep{Lin2026} is a concrete manifestation of this misalignment.

\item \textbf{Response-weighted Denoising Score Matching (RDSM).}
The RDSM framework provides a data-driven method to approximate the score function from any ellipticity catalog by integrating the shear response into a denoising score matching objective (\autoref{subsec:DSM}).
This brings the effective estimator closer to the score function ($s_g(e)=\partial\ln L(e|g)/\partial g$; \autoref{sec:shape_noise}) and reduces shape noise by $\sim\SNREL\%$ relative to FPFS+AnaCal at LSST 10-year depth while maintaining $|m|<2\times10^{-3}$ (\autoref{sec:validation}), confirming the theoretical prediction that score-aligned supervision improves shape noise without introducing bias.

\item \textbf{$w_{\rm IV}$ from conditional score identity.} The inverse variance weight $w_{\rm IV} = R(M_b)^\top / \sigma^2(M_b)$ serve as a general rescaling weight any radial family, not only the Gaussian case. 
This provides the first likelihood-based derivation of $w_{\rm IV}$ and isolates two channels for further improvement: non-radial structure in $p(\mathbf{e} \mid M_b)$ and Case~1 Fisher information from the shear response of $M_b$ (\autoref{eq:conditional-identity}).
We have further performed a preliminary test on inverse variance and score matching as shown in \autoref{sec:weight-decomp}.
\end{enumerate}

\subsection{Future Work}
\label{sec:limitations}

Several future applications follow naturally from the framework developed here.
These directions would both address the main simplifying assumptions of the present analysis and extend the method to more realistic weak-lensing scenarios:
\begin{itemize}
  \item The assumption of a known intrinsic population $f_0(I_0)$ will be validated against an external deep reference dataset such as Roman, which provides an independently measured population of high-resolution galaxy morphologies.

  \item The weak-shear (linear) approximation will be tested using image simulations with intermediate shear ($|g| \sim 0.1$--$0.3$) representative of cluster-core regimes, isolating any deviation from the linear-response prediction.

  \item The single-galaxy / no-blending assumption will be addressed by extending the score-projected training target to blended scenes and validating on blended image simulations, with appropriate detection-bias correction.

  \item The image-level score function can be more accurately approximated by combining the RDSM framework with a general image encoder that maps galaxy images into a low-dimensional spin-0 representation, enabling the score formalism to operate on compact morphology descriptors rather than just ellipticity while avoiding the full pixel space.

\end{itemize}

\begin{acknowledgments}
S. Lin and X. Liu acknowledge support from the Illinois Campus Research Board Award RB25035, U of Illinois Urbana-Champaign Center for Advanced Study, NSF grant AST-2308174, and NASA grants 80NSSC24K0219 and 80NSSC26K0333.
Xiangchong Li acknowledges support from the U.S. Department of Energy under Contract No. DE-SC0012704 and from the Laboratory Directed Research and Development (LDRD) Program at Brookhaven National Laboratory (Project No. 27992).
This work utilized resources supported by the National Science Foundation's Major Research Instrumentation program, grant \#1725729, as well as the University of Illinois at Urbana-Champaign.
This work used Delta and DeltaAI at NCSA through allocations PHY240290 and PHY250308 from the Advanced Cyberinfrastructure Coordination Ecosystem: Services \& Support (ACCESS) program, which is supported by U.S. National Science Foundation grants \#2138259, \#2138286, \#2138307, \#2137603, and \#2138296.
\end{acknowledgments}

\facilities{Rubin LSST}

\software{
astropy \citep{2013A&A...558A..33A,2018AJ....156..123A,2022ApJ...935..167A},
GalSim \citep{Rowe2015},
xlens \citep{Li2025}
}

\appendix
\section{Bayesian Framework for Shear Estimation}
\label{sec:bayesian}

\subsection{Maximum Posterior Estimation with Flat Prior}

Given $N$ measured images $\{ I_i \}_{i=1}^N$ in a region of constant shear $\mathbf{g}$, the posterior factorizes as
\begin{equation}
P(\mathbf{g} \mid \{I_i\})
\propto
P(\mathbf{g}) \prod_{i=1}^N L(I_i \mid \mathbf{g}),
\end{equation}
where $f(I\mid\mathbf{g})$ is the single-galaxy image likelihood defined in \autoref{sec:likelihood} and $P(\mathbf{g})$ the shear prior. Taking the logarithm,
\begin{align}
\ln P(\mathbf{g} \mid \{I_i\})
=
\ln P(\mathbf{g})
+ \sum_{i=1}^N \ln L(I_i \mid \mathbf{g})
+ \mathrm{const.},
\label{eq:log_prob}
\end{align}
and, under a flat prior, the maximum a posteriori (MAP) estimator coincides with the maximum likelihood estimator (MLE),
\begin{equation}
\hat{\mathbf{g}}
=
\arg\max_{\mathbf{g}}
\sum_{i=1}^N \ln L(I_i \mid \mathbf{g}).
\end{equation}
Here, we assume this likelihood satisfies the regularity conditions:
\begin{itemize}
    \item $L(I|\mathbf{g})$ is differentiable with respect to $g$ for almost all $I$;
    \item differentiation and integration can be interchanged,
    i.e., $\pard{}{g} \int L(I|\mathbf{g})\,dI = \int \pard{}{g}\,L(I|\mathbf{g})\,dI$;
    \item the Fisher information $\mathbf{F}$ is finite and positive definite.
\end{itemize}
These conditions are satisfied for the weak-lensing image model of \autoref{eq:parametrization} under mild smoothness assumptions on $f_0$, as $I$ is the signal smoothed by the PSF, which is differentiable over shear.\\
In this case, the stationarity condition for this maximum can be written as:
\begin{equation}
\sum_{i=1}^N
\frac{\partial}{\partial \mathbf{g}}
\ln f(I_i \mid \mathbf{g}) = 0,
\end{equation}
or, equivalently, that the empirical mean score must vanish at $\hat{\mathbf{g}}$,
\begin{equation}
\frac{1}{N} \sum_{i=1}^N s_{\hat{\mathbf{g}}}(I_i) = 0.
\end{equation}

In the weak-lensing regime $|\mathbf{g}| \ll 1$, we expand the log-likelihood to second order around $\mathbf{g} = \mathbf{0}$,
\begin{equation}
\ln L(I\mid\mathbf{g})
\simeq
\ln f_0(I)
+ \mathbf{g}^\top s_{\mathbf{0}}(I)
- \tfrac{1}{2}\, \mathbf{g}^\top \mathbf{F}\, \mathbf{g},
\end{equation}
where $f_0(I)\equiv L(I\mid\mathbf{0})$ is the intrinsic shape distribution and the Fisher information matrix $\mathbf{F}$ is given by \autoref{eq:fisher}. Differentiating in $\mathbf{g}$ gives the linearized score,
\begin{equation}
\pard{}{\mathbf{g}}\ln f(I\mid\mathbf{g})
\simeq
s_{\mathbf{0}}(I)
- \mathbf{F}\, \mathbf{g},
\end{equation}
and imposing the vanishing-mean-score condition above yields the closed-form linear estimator
\begin{equation}
\hat{\mathbf{g}}
=
\mathbf{F}^{-1}
\left(
\frac{1}{N}
\sum_{i=1}^N s_{\mathbf{0}}(I_i)
\right).
\label{eq:linear-mle}
\end{equation}
The optimal per-galaxy statistic for shear estimation is therefore the zero-shear score $s_{\mathbf{0}}(I_i)$, and the Fisher matrix supplies both its normalization and the estimator variance,
\begin{equation}
\mathrm{Var}(\hat{\mathbf{g}}) = \frac{1}{N}\mathbf{F}^{-1}.
\end{equation}

\subsection{Maximum Posterior Estimation with Gaussian Prior}

With a non-trivial prior, the maximum posterior would give us the following condition:
\begin{equation}
    \pard{\ln P(\mathbf{g})}{\mathbf{g}}
    + \sum_{i=1}^N
\frac{\partial}{\partial \mathbf{g}}
\ln L(\mathbf{e}_i \mid \mathbf{g}) = 0.
\end{equation}

This is usually not easy to solve. However, we can assume a Gaussian prior:
\begin{equation}
    P(\mathbf{g}) \propto \exp{-\frac{|\mathbf{g}-\mathbf{g}_0|^2}{\sigma_g^2}}.
\end{equation}
And again expanding the score function around the zero point, we have:
\begin{align}
    -2\frac{\mathbf{g}-\mathbf{g}_0}{\sigma_g^2} +
    \frac{1}{N}\sum_{i=1}^N
\left(
 s(\mathbf{e}_i|\mathbf{0})-\mathbf{F} \mathbf{g}
    \right) = 0.
\end{align}

This is a linear equation with a simple solution:
\begin{align}
    \hatbf{g} = \left(\mathbf{F}+\frac{2\,\mathbf{I}}{\sigma_g^2}\right)^{-1}
    \left(
\frac{1}{N}
\sum_{i=1}^N s(\mathbf{e}_i|\mathbf{0})+\frac{\mathbf{g}_0}{2\sigma^2_g}
\right).
\end{align}

\setcounter{equation}{0}
\renewcommand{\theequation}{B\arabic{equation}}

\section{Derivation of Score Function Formalism}

\subsection{Derivation of Response}
\label{app:r_reduction}
For response $\mathbf{R}=\pard{\mathbb{E}(\hatbf{e})}{\mathbf{g}}$,
considering the transformation of sheared and unsheared images, we have the expectation in $I_0$ space:
\begin{equation}
\begin{split}
    \pard{\mathbb{E}(\hatbf{e})}{\mathbf{g}}
    &= \pard{}{\mathbf{g}} \int \hatbf{e}[T_\mathbf{g}(I_0)] L_0(I_0) \mathrm{d}I_0 \\
    &= \int \left(\pard{}{\mathbf{g}}\hatbf{e}[T_\mathbf{g}(I_0)]\right) L_0(I_0) \mathrm{d}I_0 \\
    & = \mathbb{E}_0\left(\pard{\hatbf{e}}{\mathbf{g}} \right)
    = \mathbb{E}_0(\hatbf{R}),
\end{split}
\end{equation}
where $\mathbb{E}_0$ is used to denote that the integration has to be done in $I_0$ space.
This shows that the constant matrix $R$ coincides with the averaged response of each image.

At the same time, we also calculate the expectation in the $I$ space:
\begin{equation}
\begin{split}
    R &= \int \hatbf{e}\pard{}{\mathbf{g}}L(I|\mathbf{g})\mathrm{d}I \\
    &= \int \hatbf{e} L(I|\mathbf{g})\pard{}{\mathbf{g}} \ln L(I|\mathbf{g})\mathrm{d}I \\
    & = \int \hatbf{e} \hatbf{s}L(I|\mathbf{g}) \mathrm{d}I \\
    & = \mathbb{E}(\hatbf{e}\hatbf{s}).
\end{split}
\end{equation}

\subsection{Shape Noise in $\mathbf{e}$-Space}
\label{app:sn_espace}
To project the shape noise formula (\autoref{eq:shape_noise_angle}) into e-space, we first notice that:
\begin{equation}
\begin{split}
\mathbb{E}_I\!\left(\hat{\mathcal{E}}_a^2\right)
&= \int \hat{\mathcal{E}}_a^2(I)\, L(I \mid \mathbf{g}) \, \mathrm{d}I \\
&= \int \hat{\mathcal{E}}_a^2(I)\, L(I \mid \mathbf{g})
    \int \delta \left( \hat{\mathbf{e}} - \hat{\mathbf{e}}(I) \right)\, \mathrm{d}\hat{\mathbf{e}} \, \mathrm{d}I \\
&= \int \hat{\mathcal{E}}_a^2(\hat{\mathbf{e}})
    \left[
      \int L(I \mid \mathbf{g})
      \delta\!\left(\hat{\mathbf{e}} - \hat{\mathbf{e}}(I)\right)\, \mathrm{d}I
    \right] \mathrm{d}\hat{\mathbf{e}} \\
&= \int \hat{\mathcal{E}}_a^2(\hat{\mathbf{e}})\, L(\hat{\mathbf{e}} \mid \mathbf{g}) \, \mathrm{d}\hat{\mathbf{e}} \\
&= \mathbb{E}_{\mathbf{e}}\!\left(\hat{\mathcal{E}}_a^2\right).
\end{split}
\end{equation}

\begin{equation}
\begin{split}
\mathbb{E}_I(\hat{\mathcal{E}}_a \hat s_a)
&= \int \hat{\mathcal{E}}_a(I)\, \pard{}{g_a} L(I \mid \mathbf{g}) \, \mathrm{d}I \\
&= \int \hat{\mathcal{E}}_a(I)\, \pard{}{g_a}
    \left[
      \int \delta\!\left(\hat{\mathbf{e}} - \hat{\mathbf{e}}(I)\right)\, \mathrm{d}\hat{\mathbf{e}}
    \right]
    L(I \mid \mathbf{g}) \, \mathrm{d}I \\
&= \int \hat{\mathcal{E}}_a(\hat{\mathbf{e}})
    \left[
      \int \pard{}{g_a}
      \Big(
        L(I \mid \mathbf{g})\,
        \delta\!\left(\hat{\mathbf{e}} - \hat{\mathbf{e}}(I)\right)
      \Big)\, \mathrm{d}I
    \right]
    \mathrm{d}\hat{\mathbf{e}} \\
&= \int \hat{\mathcal{E}}_a(\hat{\mathbf{e}})\, \pard{}{g_a} L(\hat{\mathbf{e}} \mid \mathbf{g}) \, \mathrm{d}\hat{\mathbf{e}} \\
&= \int \hat{\mathcal{E}}_a(\hat{\mathbf{e}})\, L(\hat{\mathbf{e}} \mid \mathbf{g}) \, \pard{}{g_a} \ln L(\hat{\mathbf{e}} \mid \mathbf{g}) \, \mathrm{d}\hat{\mathbf{e}} \\
&= \int \hat{\mathcal{E}}_a(\hat{\mathbf{e}})\, \hat s_a(\hat{\mathbf{e}})\, L(\hat{\mathbf{e}} \mid \mathbf{g}) \, \mathrm{d}\hat{\mathbf{e}} \\
&= \mathbb{E}_{\mathbf{e}}(\hat{\mathcal{E}}_a \hat s_a).
\end{split}
\end{equation}

Combining these together, we get the shape noise in e-space:
\begin{equation}
\begin{split}
\mathrm{Var}(\hatbf{g})
&= \frac{\mathbb{E}_{\mathbf{e}}\!\left[\hat{\mathcal{E}}_a^2(\hat{\mathbf{e}})\right]}
         {\mathbb{E}_{\mathbf{e}}\!\left[\hat{\mathcal{E}}_a(\hat{\mathbf{e}})\, \hat s_a(\hat{\mathbf{e}})\right]} \\
&\ge \frac{1}{\mathbb{E}_{\mathbf{e}}\!\left[\hat s_a^2(\hat{\mathbf{e}})\right]} ,
\end{split}
\end{equation}

We note an important regularity property of the shape-space score $s_0(\mathbf{e})$ near $\mathbf{e} = 0$ here.
Since the intrinsic ellipticity distribution $L_0(\mathbf{e})$ peaks at $|\mathbf{e}| = 0$ (nearly-round galaxies are the most common), its gradient vanishes there, and consequently $s_0(\mathbf{e}) = \partial \ln L_0(\mathbf{e})/\partial \mathbf{g}|_{g=0}$ also vanishes as $|\mathbf{e}| \to 0$, as required by spin-2 symmetry.
For a well-defined linear shear response, the score must be a smooth function of $\mathbf{e}$
with a finite radial derivative $\mathrm{d}s_0/\mathrm{d}|\mathbf{e}|$ at the origin;
this implies that $s_0(\mathbf{e})$ is proportional to $\mathbf{e}$ near $\mathbf{e}=0$,
i.e., it must be linear in $\mathbf{e}$ with a finite coefficient of proportionality.
(This condition is equivalent to requiring that $\pard{s_0}{\mathbf{g}} / \pard{\mathbf{e}}{\mathbf{g}}|_{g=0}$ remain finite,
since both $\partial s_0/\partial \mathbf{g}$ and the per-object response $\partial\mathbf{e}/\partial\mathbf{g}$ are well-defined
unbiased quantities at linear order in shear.)
This has a direct consequence for the functional form of any radial transformation $\hat{\mathbf{e}}' = h(|\mathbf{e}|)\mathbf{e}$: requiring $h(0) \neq 0$ is not merely a smoothness convention but is necessary for the estimator to have a nonzero linear shear response.
Any form with $h(0) = 0$, such as $h(|\mathbf{e}|) = |\mathbf{e}|^{2n}$ ($n \geq 1$), gives an ensemble response $R \propto O(g^{2n+1})$ rather than $O(g)$ for a population dominated by round galaxies, making the weak-shear estimator degenerate. The correct general smooth $\mathrm{SO}(2)$-equivariant form is therefore $\hat{\mathbf{e}}' = h(|\mathbf{e}|)\mathbf{e}$ where $h$ is a smooth function of $|\mathbf{e}|^2$ (rather than merely of $|\mathbf{e}|$, so that the full estimator is smooth at the origin) with $h(0) = A \neq 0$, reducing near the origin to $A\mathbf{e}$ with finite response.
The choice between conditioning on or marginalizing over auxiliary observables such as $M_{00}$ is treated separately in \autoref{sec:weight-hierarchy}; this distinction has direct empirical consequences for the calibration of RDSM estimators in \autoref{sec:weight-decomp}.

\subsection{Parametrization of Score Function in $\mathbf{e}$-Space}
\label{app:e_space_param}
By definition, the score function $s(\mathbf{e}|\mathbf{g})$ at zero shear can be expressed by:
\begin{equation}
\begin{split}
s_\mathbf{0}(\mathbf{e})
&= \frac{1}{L_0(\mathbf{e})}\pard{}{\mathbf{g}} \int \mathrm{d}I L(I|\mathbf{0}) \delta(\mathbf{e}-\hatbf{e}(I)) \\
&= \frac{1}{L_0(\mathbf{e})}
\int \mathrm{d}I \pard{L(I|\mathbf{0})}{\mathbf{g}} \delta(\mathbf{e}-\hatbf{e}(I)) \\
&= \frac{1}{L_0(\mathbf{e})}
\int \mathrm{d}I \pard{}{\mathbf{g}}
\left[
D(I, \mathbf{g})\, L(I|\mathbf{0})
\right]
\delta(\mathbf{e}-\hatbf{e}(I)) \\
&= \frac{1}{L_0(\mathbf{e})}\int \mathrm{d}I L(I|\mathbf{0})
\delta(\mathbf{e}-\hatbf{e}(I)) \pard{}{\mathbf{g}}D(I, \mathbf{g}) \\
&+\frac{1}{L_0(\mathbf{e})} \int \mathrm{d}I
\pard{}{\mathbf{g}} L(I|\mathbf{0}) \delta(\mathbf{e}-\hatbf{e}(I)).
\end{split}
\end{equation}
For the second term, we integrate by parts:
\begin{equation}
\begin{split}
   & \frac{1}{L_0(\mathbf{e})} \int \mathrm{d}I
\pard{}{\mathbf{g}} L(I|\mathbf{0}) \delta(\mathbf{e}-\hatbf{e}(I)) \\
&=- \frac{1}{L_0(\mathbf{e})} \int  \mathrm{d}I
L(I|\mathbf{0})\pard{I}{\mathbf{g}} \pard{}{I}\delta(\mathbf{e}-\hatbf{e}(I)) \\
&= - \frac{1}{L_0(\mathbf{e})} \int  \mathrm{d}I
L(I|\mathbf{0})\pard{I}{\mathbf{g}}
\pard{\mathbf{e}}{I}\pard{}{\mathbf{e}}\delta(\mathbf{e}-\hatbf{e}(I)) \\
&= - \frac{1}{L_0(\mathbf{e})} \int  \mathrm{d}I
L(I|\mathbf{0})\pard{I}{\mathbf{g}}
\left(\pard{\mathbf{e}}{I}\right)^\mathrm{T}
\pard{}{\mathbf{e}}\delta(\mathbf{e}-\hatbf{e}(I)) \\
&= - \frac{1}{L_0(\mathbf{e})} \int  \mathrm{d}I
L(I|\mathbf{0})\mathbf{R}(I)
\pard{}{\mathbf{e}}\delta(\mathbf{e}-\hatbf{e}(I)) \\
&= - \frac{1}{L_0(\mathbf{e})} \pard{}{\mathbf{e}}\int  \mathrm{d}I
L(I|\mathbf{0})\mathbf{R}(I)
\delta(\mathbf{e}-\hatbf{e}(I)).
\end{split}
\end{equation}

We define the average ellipticity response and detection response:
\begin{align}
    \bar{\mathbf{R}}_0(\mathbf{e}) =
    \frac{1}{L_0(\mathbf{e})} \int  \mathrm{d}I
    L(I|\mathbf{0})\mathbf{R}(I)
    \delta(\mathbf{e}-\hatbf{e}(I))
\end{align}

\begin{align}
    \bar{D}(\mathbf{e},\mathbf{g}) =
    \frac{1}{L_0(\mathbf{e})} \int  \mathrm{d}I
    L(I|\mathbf{0})D(I|\mathbf{g})
    \delta(\mathbf{e}-\hatbf{e}(I))
\end{align}
Finally, we can express the score function as:
\begin{equation}
\begin{split}
    s_\mathbf{0}(\mathbf{e})
&= \pard{}{\mathbf{g}} (L_0(\mathbf{e})\bar{D}(\mathbf{e},\mathbf{0}))
- \frac{1}{L_0(\mathbf{e})} \pard{}{\mathbf{e}}\cdot[L_0(\mathbf{e})\bar{\mathbf{R}}_0(\mathbf{e})] \\
&= \pard{}{\mathbf{g}} (L_0(\mathbf{e})\bar{D}(\mathbf{e},\mathbf{0}))
-  \pard{\ln L_0(\mathbf{e})}{\mathbf{e}}\cdot\bar{\mathbf{R}}_0(\mathbf{e})
-  \pard{}{\mathbf{e}}\cdot \bar{\mathbf{R}}_0(\mathbf{e}).
\end{split}
\end{equation}

\subsection{Derivation of RDSM Target}
\label{app:rdsm_target}
With this, we first need the smoothed likelihood:
\begin{align}
L(\tilde{\mathbf M} \mid \mathbf g)
=
\int d\mathbf M \;
L(\mathbf M \mid \mathbf g)\,
\mathcal N(\tilde{\mathbf M} \mid \mathbf M, \sigma^2 \mathbf I).
\end{align}
Substituting the expression for $L(\mathbf M \mid \mathbf g)$ and integrating over $\mathbf M$ gives:
\begin{align}
L(\tilde{\mathbf M} \mid \mathbf g)
=
\int dI_0 \;
L_0(I_0)\,
\mathcal N\!\Big(
\tilde{\mathbf M}
\;\Big|\;
\hatbf{M}(T_\mathbf{g}(I_0)) \Big)
\end{align}
For a Gaussian distribution,
\begin{align}
\nabla_{\mathbf g}
\ln
\mathcal N\Big(\tilde{\mathbf M}
\Big| \hatbf{M}(T_\mathbf{g}(I_0)), \sigma^2 \mathbf I \Big)
=
\frac{\mathbf R(I_0)^{\!\top}
\left(
\tilde{\mathbf M}
-
\hatbf{M}(T_\mathbf{g}(I_0))
\right)}
{\sigma^2}.
\end{align}

Therefore,
\begin{equation}
    \begin{aligned}
        &\nabla_{\mathbf g}
        \ln L(\tilde{\mathbf M} \mid \mathbf g) \\
        &= \frac{1}{L(\tilde{\mathbf M}|\mathbf{g})} \int
        \mathrm{d}I_0 L_0(I_0)
        \frac{\mathbf R(I_0)^{\!\top} \left( \tilde{\mathbf M} - \hatbf{M}(T_\mathbf{g}(I_0)) \right)}
        {\sigma^2} \mathcal N\!\Big(
        \tilde{\mathbf M}
        \;\Big|\;
        \hatbf{M}(T_\mathbf{g}(I_0)) \Big)
        \\
        &=\mathbb E_{I_0 \mid \tilde{\mathbf M}, \mathbf g}
        \left[
        \frac{\mathbf R(I_0)^{\!\top}
        \left(
        \tilde{\mathbf M}-
        \hatbf{M}(T_\mathbf{g}(I_0))
        \right)
        }
        {\sigma^2}
        \right].
    \end{aligned}
\end{equation}

Notice that
$\mathcal N\!\Big(\tilde{\mathbf M} \Big| \hatbf{M}(T_\mathbf{g}(I_0)) \Big)
= L(\tilde{\mathbf M}|I_0,\mathbf{g})
$; by Bayes' theorem, we have:
\begin{align}
    L(I_0|\tilde{\mathbf M},\mathbf{g}) =
    \frac{L_0(I_0) \mathcal{N}\Big(\tilde{\mathbf M} \Big| \hatbf{M}(T_\mathbf{g}(I_0)) \Big)}
    {L(\tilde{\mathbf M}|\mathbf{g})}.
\end{align}
As a result:
\begin{align}
    \nabla_{\mathbf g}
\ln L(\tilde{\mathbf M} \mid \mathbf g)=
\mathbb E_{I_0 \mid \tilde{\mathbf M}, \mathbf g}
\left[
\frac{\mathbf R(I_0)^{\!\top}
\left(
\tilde{\mathbf M}-
\hatbf{M}(T_\mathbf{g}(I_0))
\right)
}
{\sigma^2}
\right].
\end{align}

Evaluating at $\mathbf g = \mathbf 0$ gives
\begin{align}
\mathbf s_\mathbf{0}(\tilde{\mathbf M})
=
\mathbb E_{I_0 \mid \tilde{\mathbf M}}
\left[
\frac{\mathbf R(I_0)^{\!\top}
\left(
\tilde{\mathbf M}
-
\hat{\mathbf M}(I_0)
\right)}
{\sigma^2}
\right].
\end{align}

Since the noisy moments are generated as
\begin{align}
\tilde{\mathbf M}
=
\hat{\mathbf M}(I_0)
+
\boldsymbol\epsilon,
\end{align}
the score becomes
\begin{align}
\mathbf s(\tilde{\mathbf M})
=
\mathbb E
\left[
\frac{\mathbf R(I_0)^{\!\top}
\boldsymbol\epsilon}
{\sigma^2}
\;\middle|\;
\tilde{\mathbf M}
\right].
\end{align}

Such an expectation can be realized by
training a network $s_\theta(\tilde{\mathbf M})$ with the following MSE loss:
\begin{align}
\mathcal L
=
\mathbb E
\left\|
s_\theta(\tilde{\mathbf M})
-
\frac{\mathbf R(I_0)^{\!\top}\boldsymbol\epsilon}{\sigma^2}
\right\|^2.
\end{align}

\subsection{Score Function Decomposition}
\label{app:decomposition}

Under the weak-lensing limit, we can consider a linear parametrization for the image $I$ as mentioned in \autoref{eq:parametrization}:
\begin{align}
    I = T_\mathbf{g}(I_0) = I_0+\pard{I}{g}g.
\end{align}
With this parametrization, we further decompose the likelihood by separating the intrinsic image distribution from the shear-dependent selection and detection:
\begin{equation}
  L(I \mid g) = D(I, g)\, L_0(I_0)  \exp\!\left[-g\,\mathrm{tr}\!\left(\frac{\partial^2 T_\mathbf{g}^{-1}(I)}{\partial I\,\partial g}\right)\right],
  \label{eq:L_decomposed}
\end{equation}
where $D(I, g)$ is the \emph{detection weight} (the probability that a galaxy
with image $I$ under shear $g$ is included in the sample). For a hard magnitude
cut at threshold $\alpha_0$, $D = \Theta(\alpha(I,g) - \alpha_0)$ is a
Heaviside step function in the measured flux or size $\alpha$, whose shear
dependence introduces the detection-selection bias familiar from
\textsc{Metadetection} \citep{Sheldon2020} and AnaCal \citep{Li2023}. More
generally, $D$ can encode any selection that depends on shear-dependent
observables, including blending flags, S/N cuts, and photometric redshift
requirements.
The score function under this parametrization
is then:
\begin{equation}
\begin{split}
& \hat s(I \mid \mathbf{g}) \\
&= \pard{}{\mathbf{g}}
   \Big[
     \ln L_0\!\left(I-\mathbf{g}\,\Big.\pard{I}{\mathbf{g}}\Big|_{I}\right)
     + \ln D(I, \mathbf{g})
     - \mathbf{g}\,
       \mathrm{tr}\!\left(\frac{\partial^2 T_\mathbf{g}^{-1}(I)}{\partial I\,\partial g}\right)
   \Big] \\
&=
\pard{}{\mathbf{g}}
\ln L_0\!\left(I-\mathbf{g}\,\Big.\pard{I}{\mathbf{g}}\Big|_{I}\right)
+ \pard{}{\mathbf{g}}\ln D(I, \mathbf{g})
- \mathrm{tr}\!\left(\frac{\partial^2 T_\mathbf{g}^{-1}(I)}{\partial I\,\partial g}\right) .\\
&= \pard{}{\mathbf{g}}\ln D(I, \mathbf{g})
- \partial_I \ln L_0(I_0) \cdot
\Big.\pard{I}{\mathbf{g}}\Big|_{I}
- \mathrm{tr}\!\left(\frac{\partial^2 T_\mathbf{g}^{-1}(I)}{\partial I\,\partial g}\right).
\label{eq:score_decomp_image}
\end{split}
\end{equation}

The above image-space score can be re-expressed as a function of ellipticity $\mathbf{e}$ alone,
with an averaged response matrix $\bar{\mathbf{R}}_0(\mathbf{e})$ and effective detection weight $\bar{D}(\mathbf{e},\mathbf{0})$ that includes contributions from both the intrinsic shape response and the detection weight response.
Readers can find details of the derivation in \autoref{app:e_space_param}.
The whole score is expressed as:
\begin{equation}
\begin{split}
    s_\mathbf{0}(\mathbf{e})
&= \frac{1}{L_0(\mathbf{e})}\pard{}{\mathbf{g}} (L_0(\mathbf{e})\bar{D}(\mathbf{e},\mathbf{0}))
- \frac{1}{L_0(\mathbf{e})} \pard{}{\mathbf{e}}\cdot[L_0(\mathbf{e})\bar{\mathbf{R}}_0(\mathbf{e})] \\
&= \frac{1}{L_0(\mathbf{e})}\pard{}{\mathbf{g}} (L_0(\mathbf{e})\bar{D}(\mathbf{e},\mathbf{0}))
-  \pard{\ln L_0(\mathbf{e})}{\mathbf{e}}\cdot\bar{\mathbf{R}}_0(\mathbf{e})\\
&-  \pard{}{\mathbf{e}}\cdot \bar{\mathbf{R}}_0(\mathbf{e}).
\label{eq:score_decomp}
\end{split}
\end{equation}
The score function under this parametrization is separated into three terms:
\begin{itemize}
        \item $\displaystyle\frac{1}{L_0(\mathbf{e})}\pard{}{\mathbf{g}} (L_0(\mathbf{e})\bar{D}(\mathbf{e},\mathbf{0}))$: the shear response of the detection weight.
    \item $\displaystyle - \pard{\ln L_0(\mathbf{e})}{\mathbf{e}}\cdot\bar{\mathbf{R}}_0(\mathbf{e})$: the projection of the gradient of the likelihood onto the shear direction.
    \item $\displaystyle - \pard{}{\mathbf{e}}\cdot \bar{\mathbf{R}}_0(\mathbf{e})$: the divergence of the response matrix, coming from the change of volume in parameter space under the parametrization.
\end{itemize}
The above formula works not only for ellipticity but also for any variables based on the image.

As a special case, the classical complex distortion $\tilde{\epsilon}$ transforms via the M\"{o}bius form $\tilde{\epsilon}' = (\tilde{\epsilon} + \tilde{g})/(1 + \tilde{g}\cdot\tilde{\epsilon})$ \citep[][Eq.~3-31]{Bernstein2002}. Projecting onto the real 2-vector representation, the sheared distortion can be written as
\begin{equation}
    \mathbf{e}(\mathbf{g}) = \mathbf{e} + \mathbf{g} - \mathbf{e}\,(\mathbf{g}\cdot\mathbf{e}) + O(g^2),
\label{eq:linear-shear-model}
\end{equation}

The per-image response matrix follows immediately from \autoref{eq:linear-shear-model}:
\begin{equation}
    \hat{R}_{ab}(\mathbf{e}) \equiv \frac{\partial e_a(\mathbf{g})}{\partial g_b}\bigg|_{\mathbf{g}=0} = \delta_{ab} - e_a e_b.
\label{eq:response-linear}
\end{equation}
And the divergence of this response is:

\begin{equation}
    \frac{\partial \bar{R}_{ba}}{\partial e_b} = \frac{\partial}{\partial e_b}(\delta_{ba} - e_b e_a) = -3 e_a .
\end{equation}

Assuming no detection-selection bias ($D = 1$) and an isotropic shape distribution $L_0(|\mathbf{e}|)$,
the averaged response $\bar{R}_{ab}$ inherited from \autoref{eq:response-linear} is simply $\hat{R}_{ab}(\mathbf{e})$, and the score decomposition of \autoref{eq:score_decomp} reduces to
\begin{align}
    s_{0,a}(\mathbf{e})
    &= -\frac{\partial \ln L_0(|\mathbf{e}|)}{\partial e_b}\,\bar{R}_{ba}
       - \frac{\partial}{\partial e_b}\bar{R}_{ba} \\
    &= -\frac{\partial \ln L_0(|\mathbf{e}|)}{\partial |\mathbf{e}|}
    \frac{(1-|\mathbf{e}|^2)e_a}{|\mathbf{e}|}
       + 3 e_a,
\label{eq:score-special}
\end{align}

\setcounter{equation}{0}
\renewcommand{\theequation}{C\arabic{equation}}

\section{Full Shapelet Mode Transformations Under Shear}
\label{app:FPFS}

For completeness, we list the first-order shear transformations of the relevant FPFS
shapelet modes \citep{Refregier2003, Massey2005}:
\begin{equation}
\begin{split}
  M'_{00} &=
    M_{00} -\sqrt{2}\,\Big(g_{1}\,M_{22c}+g_{2}\,M_{22s}\Big), \\
  M'_{20} &=
   M_{20} -\sqrt{6}\,\Big(g_{1}\,M_{42c}+g_{2}\,M_{42s}\Big), \\
  M'_{22c} &=
    M_{22c} +\frac{\sqrt{2}}{2}\,(M_{00}-M_{40})\,g_{1}
    -\sqrt{3}\,\Big(g_{1}\,M_{44c}+g_{2}\,M_{44s}\Big), \\
  M'_{22s} &=
    M_{22s}
    +\frac{\sqrt{2}}{2}\,(M_{00}-M_{40})\,g_{2}
    -\sqrt{3}\,\Big(g_{1}\,M_{44s}-g_{2}\,M_{44c}\Big), \\
  M'_{11c} &=
     M_{11c}
     +\frac{\sqrt{2}}{2}\,\Big(g_{1}\,M_{31c}+g_{2}\,M_{31s}\Big)
     -\frac{\sqrt{6}}{2}\,\Big(g_{1}\,M_{33c}+g_{2}\,M_{33s}\Big), \\
  M'_{11s} &=
    M_{11s}
     +\frac{\sqrt{2}}{2}\,\Big(g_{2}\,M_{31c}-g_{1}\,M_{31s}\Big)
     -\frac{\sqrt{6}}{2}\,\Big(g_{1}\,M_{33s}-g_{2}\,M_{33c}\Big).
\end{split}
\end{equation}
These transformations are used in \autoref{sec:basis_projection} to derive the
optimal score estimator in FPFS moment space.

We define the FPFS ellipticity components as
\begin{equation}
  e_1 \equiv \frac{M_{22c}}{M_{00}+C}, \qquad
  e_2 \equiv \frac{M_{22s}}{M_{00}+C}\,.
  \label{eq:fpfs_ellipticity}
\end{equation}

Suppose the joint distribution of ellipticity and flux moment $M_{00} \equiv
M_{00}$ at zero shear takes the Gaussian form
\begin{equation}
  P(e_1, e_2, M_{00})
  = \frac{1}{2\pi\,\sigma_g^2(M_{00})}\,
    \exp\!\left(-\frac{e_1^2+e_2^2}{2\,\sigma_g^2(M_{00})}\right)
    \,P_0(M_{00}),
  \label{eq:gaussian-marginal}
\end{equation}
where $\sigma_g(M_{00})$ is the shape dispersion that depends on galaxy
brightness and can be estimated by binning galaxies in $M_{00}$, and
$P_0(M_{00})$ is the marginal distribution of $M_{00}$.

The log-likelihood for a single galaxy is
\begin{equation}
  \mathcal{L} \equiv \ln P
  = -\frac{e_1^2+e_2^2}{2\,\sigma_g^2(M_{00})}
    - \ln\!\big(2\pi\,\sigma_g^2(M_{00})\big)
    + \ln P_0(M_{00}).
  \label{eq:log_likelihood_gaussian}
\end{equation}

The score function (the derivative of the log-likelihood with respect to the
shear component $g_a$) is obtained via the chain rule:
\begin{equation}
  \frac{\partial\mathcal{L}}{\partial g_a}
  = \frac{\partial\mathcal{L}}{\partial e_1}\,\frac{\partial e_1}{\partial g_a}
  + \frac{\partial\mathcal{L}}{\partial e_2}\,\frac{\partial e_2}{\partial g_a}
  + \frac{\partial\mathcal{L}}{\partial M_{00}}\,\frac{\partial M_{00}}{\partial g_a}.
  \label{eq:score_chain_rule}
\end{equation}

From \autoref{eq:log_likelihood_gaussian}, the partial derivatives with respect
to the ellipticity components are
\begin{equation}
  \frac{\partial\mathcal{L}}{\partial e_\alpha}
  = -\frac{e_\alpha}{\sigma_g^2(M_{00})},
  \qquad \alpha = 1,2.
  \label{eq:dL_de}
\end{equation}
The partial derivative with respect to $M_{00}$ is
\begin{equation}
  \frac{\partial\mathcal{L}}{\partial M_{00}}
  = \frac{(e_1^2+e_2^2)}{2\,\sigma_g^4}\,\frac{\mathrm{d}\sigma_g^2}{\mathrm{d}M_{00}}
    - \frac{1}{\sigma_g^2}\,\frac{\mathrm{d}\sigma_g^2}{\mathrm{d}M_{00}}
    + \frac{\mathrm{d}\ln P_0}{\mathrm{d}M_{00}}.
  \label{eq:dL_dm00}
\end{equation}

We model the average shear response of $e_1$, $e_2$ as a function of $M_{00}$
\begin{equation}
\begin{split}
  \frac{\partial e_1}{\partial g_1} &= R(M_{00}) \\
  \frac{\partial e_2}{\partial g_2} &= R(M_{00})
\end{split}
\label{eq:de_dg_explicit}
\end{equation}
and
\begin{equation}
\frac{\partial M_{00}}{\partial g_a} = -\sqrt{2}\,M_{22a},
\qquad a = 1\,(c),\; 2\,(s).
\label{eq:m00-shear-response}
\end{equation}

Assembling \autoref{eq:score_chain_rule} using \autoref{eq:dL_de},
\autoref{eq:dL_dm00}, \autoref{eq:de_dg_explicit}, and \autoref{eq:m00-shear-response},
the score for the $g_1$ component is
\begin{equation}
s_1
\equiv \frac{\partial\mathcal{L}}{\partial g_1}
    = -e_1\left(\frac{R(M_{00})}{\sigma_g^2} + \sqrt{2} \Phi(M_{00}) \right),
\label{eq:s1_gaussian}
\end{equation}
where
\begin{equation}
\Phi(M_{00}) \equiv
    \left(\frac{|e|^2}{2\,\sigma_g^4}\,\frac{\mathrm{d}\sigma_g^2}{\mathrm{d}M_{00}}
    - \frac{1}{\sigma_g^2}\,\frac{\mathrm{d}\sigma_g^2}{\mathrm{d}M_{00}}
    + \frac{\mathrm{d}\ln P_0}{\mathrm{d}M_{00}} \right) \left(M_{00} + C\right)
\label{eq:Phi_m00}
\end{equation}
encapsulates the $M_{00}$-dependent contribution, with $|e|^2 = e_1^2+e_2^2$.
An analogous expression holds for $s_2$ with $e_1 \leftrightarrow e_2$:
\begin{equation}
s_a = -e_a\left(\frac{R(M_{00})}{\sigma_g^2}
+ \sqrt{2}\,\Phi(M_{00})\right),
\label{eq:gaussian-marginal-score}
\end{equation}
where the first term is the optimal ``inverse-variance'' weight, and the second
term captures the $M_{00}$-dependent correction arising from the brightness
dependence of the shape dispersion $\sigma_g(M_{00})$ and the flux distribution
$P_0(M_{00})$.

We now isolate the content of each term to construct a hierarchy of closed-form weights that bracket the full Gaussian-marginal optimum.

\subsection{Ensemble-Average Decomposition of \texorpdfstring{$\Phi$}{Phi}}
\label{app:phi-decomp}

We split $\Phi(M_{00})$ into a shape-dependent piece and a flux-only piece:
\begin{align}
\Phi_{\rm shape}(|e|, M_{00}) &\equiv \left[\frac{|e|^{2}}{2\sigma_g^4} - \frac{1}{\sigma_g^2}\right]\frac{d\sigma_g^2}{dM_{00}}\,(M_{00}+C), \label{eq:phi-shape}\\
\Phi_{\rm flux}(M_{00}) &\equiv (M_{00}+C)\,\frac{d\ln P_0}{dM_{00}}. \label{eq:phi-flux}
\end{align}
For the circular Gaussian of \autoref{eq:gaussian-marginal}, the conditional expectation of $|e|^{2}$ at fixed $M_{00}$ is $\langle|e|^{2}\rangle_{M_{00}} = 2\sigma_g^{2}(M_{00})$, so
\begin{equation}
\langle\Phi_{\rm shape}\rangle_{|e|\mid M_{00}} = 0, \qquad \langle\Phi\rangle_{|e|\mid M_{00}} = \Phi_{\rm flux}(M_{00}).
\label{eq:phi-conditional-mean}
\end{equation}
The shape-dependent piece carries per-galaxy $|e|^{2}$ information and averages to zero at fixed $M_{00}$; only the flux-gradient piece survives a conditional average over $|e|$.

\subsection{Hierarchy of Closed-Form Weights}
\label{app:weight-hierarchy}

Combining \autoref{eq:gaussian-marginal-score} with the decomposition of \autoref{eq:phi-conditional-mean}, we define four closed-form weight schemes that use progressively more of the Fisher information content of the Gaussian marginal model. The shear estimator for each is $\hat{g}_a = \sum_i w_i\,e_{a,i} / \sum_i w_i R(m_{00,i})$, with weights:
\begin{align}
w_{\rm IV}(M_{00}) &= \frac{R(M_{00})}{\sigma_g^{2}(M_{00})}, \label{eq:w-IV}\\
w_{\rm J}(M_{00}) &= \frac{R(M_{00})}{\sigma_g^{2}(M_{00})} + \sqrt{2}\,(M_{00}+C)\left[\frac{d\ln P_0}{dM_{00}} - \frac{1}{2\sigma_g^{2}}\frac{d\sigma_g^{2}}{dM_{00}}\right], \label{eq:w-J}\\
w_{\rm flux}(M_{00}) &= \frac{R(M_{00})}{\sigma_g^{2}(M_{00})} + \sqrt{2}\,\Phi_{\rm flux}(M_{00}), \label{eq:w-flux}\\
w_{\rm full}(|e|, M_{00}) &= \frac{R(M_{00})}{\sigma_g^{2}(M_{00})} + \sqrt{2}\left[\Phi_{\rm shape}(|e|, M_{00}) + \Phi_{\rm flux}(M_{00})\right]. \label{eq:w-full}
\end{align}
$w_{\rm IV}$ is the standard inverse-variance weight that is widely used but lacks a derivation from the full likelihood. $w_{\rm J}$ is the simple-Jacobian heuristic obtained by dropping the $|e|^{2}$-dependent term and retaining the per-component log-Jacobian $-\tfrac{1}{2}d\ln\sigma_g^{2}/dM_{00}$; it is closed form in $M_{00}$ only. $w_{\rm flux}$ is the rigorous 2D conditional mean of $w_{\rm full}$ given $M_{00}$ and is also closed form in $M_{00}$. $w_{\rm full}$ is the full Gaussian-marginal score estimator; it uses per-galaxy $|e|^{2}$ and is the upper bound within the Gaussian-marginal model. The Gaussian-marginal Cram\'{e}r--Rao bound is saturated by $w_{\rm full}$ but not by $w_{\rm J}$, $w_{\rm flux}$, or $w_{\rm IV}$.
In \autoref{sec:weight-hierarchy} and \autoref{sec:weight-decomp}, we have shown results about $w_{\rm IV}$.
Implementation of $w_{\rm J}$, $w_{\rm flux}$, and $w_{\rm full}$ is not straightforward in practice, as they require analytical form of $\sigma_g^{2}(M_{00})$ and $P_0(M_{00})$.

Both $\sigma_g^{2}(M_{00})$ and $P_0(M_{00})$ can be estimated from an unlensed deep reference catalog. To avoid noisy numerical second derivatives, we recommend smooth parametric fits: a power law $\sigma_g^{2}(M_{00}+C) = A(M_{00}+C)^{\alpha}$ and a cubic spline or log-polynomial for $\ln P_0(M_{00}+C)$. Derivatives are then analytic in the fit coefficients, and the shift $C > -\min(M_{00})$ ensures positivity. None of $w_{\rm IV}$, $w_{\rm J}$, $w_{\rm flux}$ requires image simulations or network training.

\subsection{Score for a General Radial Conditional Family}
\label{app:radial-family-score}

For the conditional of \autoref{eq:radial-family}, with $u = \|e - R(m)g\|^{2}/[2 \sigma^{2}(m)]$,
\begin{equation*}
\log p(e \mid g, m) = -\log Z(m) + \log q_m(u).
\end{equation*}
Since $Z(m)$ does not depend on $g$, the score is
\begin{equation*}
\nabla_g \log p(e \mid g, m) = \frac{q_m'(u)}{q_m(u)} \nabla_g u = -\frac{q_m'(u)}{q_m(u)} \cdot \frac{R^{\top}(m)\, [e - R(m)\, g]}{\sigma^{2}(m)}.
\end{equation*}
Evaluated at $g = 0$, this gives \autoref{eq:cond-score}. The direction depends on $(m, e)$ only through $R^{\top}(m)\, e / \sigma^{2}(m)$; the radial profile $q_m$ enters only through the scalar $q_m'(u)/q_m(u)$. Hence the inverse-variance weight $w_{\rm IV} = R(m)/\sigma^{2}(m)$ is the optimal linear weight of the form $w(m)$ for any radial family, not only Gaussian.

\setcounter{equation}{0}
\renewcommand{\theequation}{D\arabic{equation}}

\section{Projection onto Basis Functions: Worked Example in FPFS Moment Space}\label{sec:basis_projection}

A more numerically stable alternative to learning the full score function is to project it onto a finite set of basis functions $\{\phi_k(I)\}$. Using the identity
\begin{equation}
  \mathbb{E}[\phi_k(I)\,\hat{s}_a(I)] = \frac{\partial}{\partial g_a}\mathbb{E}[\phi_k(I)],
  \label{eq:moment_condition}
\end{equation}
one can define a loss that enforces these moment conditions for a model $s_\theta(I)$; the right-hand side is precisely the quantity computed by AnaCal for any differentiable observable $\phi_k$. The learned score is then optimal within the subspace spanned by $\{\phi_k\}$, and \autoref{eq:moment_condition} can be viewed as the closed-form solution to the projection problem on a restricted function space.

As a concrete example, we work in the space of FPFS moments \citep{FPFS2018,FPFS2022}, following the shapelet mode transformation under shear \citep{Refregier2003,Massey2005}:
\begin{equation}
\begin{split}
  M'_{00} &= M_{00} - \sqrt{2}\left(g_1 M_{22c} + g_2 M_{22s}\right), \label{eq:M00} \\
  M'_{22c} &= M_{22c} + \frac{\sqrt{2}}{2}(M_{00} - M_{40})\,g_1
    - \sqrt{3}\left(g_1 M_{44c} + g_2 M_{44s}\right),
\end{split}
\end{equation}
and analogous expressions for $M'_{20}$, $M'_{22s}$, $M'_{11c}$, $M'_{11s}$ (see \autoref{app:FPFS} for the full set). Retaining only the dominant shear-dependent modes $M_{00}$ and $m_{22c}$, the joint probability transforms as:
\begin{equation}
\begin{split}
 & P'(m'_{00}, m'_{22c} \mid g_1) \\
 & = \frac{1}{1+g_1^2}
    P\!\left(m'_{00} + \sqrt{2}\,g_1\,m'_{22c},\;
            m'_{22c} - \tfrac{\sqrt{2}}{2}(m'_{00} - \bar{m}_{40})\right),
  \label{eq:P_transformed}
\end{split}
\end{equation}
where we have approximated $m_{40} \approx \bar{m}_{40}$ (constant across the sample). Differentiating with respect to $g_1$ at $g_1 = 0$:
\begin{equation}
  \frac{\partial P'(m)}{\partial g_1}\bigg|_{g_1=0}
  = \sqrt{2}\,\frac{\partial P}{\partial M_{00}}\,m_{22c}
  - \frac{\sqrt{2}}{2}\,\frac{\partial P}{\partial m_{22c}}(M_{00} - \bar{m}_{40}).
  \label{eq:dPdg}
\end{equation}
The optimal score in this two-dimensional moment space is therefore:

 \begin{equation}
    \begin{aligned}
         s_1
     & = \sum_j \frac{\partial P'\left(\vm^{(j)}\right)}{\partial g_1} \Big/ P'\left(\vm^{(j)}\right) \\
     & = \sum_j \left(\sqrt{2}\frac{\partial \ln\Big(P'\left(\vm^{(j)}\right)\Big)}{\partial M_{00}} m_{22c}
         - \frac{\sqrt{2}}{2} \frac{\partial \ln\Big(P'\left(\vm^{(j)}\right)\Big)}{\partial m_{22c}} (M_{00} - \bar{m}_{40})
         \right)
    \end{aligned}
\label{eq:s1_fpfs}
 \end{equation}

which is the maximum-likelihood score estimator in FPFS space. In practice, $\partial\ln P/\partial M_{00}$ and $\partial\ln P/\partial m_{22c}$ can be estimated nonparametrically from a large unlensed galaxy sample (e.g., from a deep reference field), making this approach self-calibrating. This worked example demonstrates that the score-based framework is directly applicable to catalog-level data without requiring pixel-level model fitting.

\autoref{eq:moment_condition} also holds when $\phi_k$ is defined in the low-dimensional ellipticity or FPFS moment space rather than in pixel space; in that reduced space, a representative functional basis is much easier to estimate from data.

\bibliographystyle{aasjournalv7}
\bibliography{main}
\end{document}